\documentclass[12pt,preprint]{aastex}

\lefthead{Laws et al.}
\righthead{New Abundance Analyses of 30 Systems}
\slugcomment{Submitted to the Astronomical Journal}
\received{}
\accepted{}
\begin{document}

\title{Parent Stars of Extrasolar Planets VII: New Abundance 
Analyses of 30 Systems}

\author{Chris Laws\altaffilmark{1}, 
Guillermo Gonzalez\altaffilmark{2}, Kyle M. Walker\altaffilmark{3},
Sudhi Tyagi\altaffilmark{1}, Jeremy Dodsworth\altaffilmark{1},
Keely Snider\altaffilmark{1}, Nicholas B. Suntzeff\altaffilmark{4}}

\altaffiltext{1}{University of Washington, Astronomy Department, Box 351580, 
Seattle, WA 98195 laws@astro.washington.edu}
\altaffiltext{2}{Department of Physics and Astronomy, Iowa State University, 
Ames, IA 50011, gonzog@iastate.edu}
\altaffiltext{3}{Department of Astronomy, The Ohio State University, Columbus, OH 43210}
\altaffiltext{4}{Cerro Tololo Inter-American Observatory, NOAO, Casilla 603, La Serena,
Chile, nsuntzeff@noao.edu}

\begin{abstract}

The results of new spectroscopic analyses of 30 stars with giant
planet and/or brown dwarf companions are presented. Values for $T_{\rm eff}$
and [Fe/H] are used in conjunction with {\it Hipparcos} data and Padova
isochrones to derive masses, ages, and theoretical surface gravities.
These new data are combined with spectroscopic and photometric metallicity
estimates of other stars harboring planets and published samples of F, G, and
K dwarfs to compare several subsets of planet bearing stars with similarly
well-constrained control groups. The distribution of [Fe/H] values continues
the trend uncovered in previous studies in that stars hosting planetary 
companions have a higher mean value than otherwise similar nearby stars.
We also investigate the relationship between stellar mass and the presence 
of giant planets and find statistically marginal but suggestive evidence of 
a decrease in the incidence of radial velocity companions orbiting relatively 
less massive stars. If confirmed with larger samples, this would represent a 
critical constraint to both planetary formation models as well as to estimates 
of the distribution of planetary systems in our galaxy.

\end{abstract}


\section{Introduction}
\label{intro}

We have steadily reported the results of our spectroscopic analyses of
stars-with-planets (hereafter, SWPs) in our series of studies on 
this topic (Gonzalez 1997, Paper I; Gonzalez 1998, Paper II; Gonzalez \&
Vanture 1998, Paper III; and Gonzalez et al. 1999, Paper IV; Gonzalez \&
Laws 2000, Paper V; Gonzalez et al. 2001, Paper VI), and continue this effort 
here with our most recent findings. Other similar studies include Fuhrmann
et al. (1998), Santos et al. (2000, 2001a, 2002), Takeda et al. (2001), and Zhao
et al. (2002). Such intensive, detailed, and persistent spectroscopic programs have
been demanded by the observed correlations between chemical abundances and other
characteristics among SWPs (c.f. Paper VI, Santos et al. 2002, Gonzalez 2003).
By providing well-constrained stellar parameters for a homogeneously analyzed
cohort of SWPs, we hope to provide further insight into these trends and to 
support research into the underlying physics which gives rise to them.

In Section 2 we discuss our program of observations, and in Section 3 we
present a brief review of our analysis method and new results for 25 SWPs,
along with updated parameters for 5 previously analyzed systems. Additionally,
we report on HD\,202206, which has been reported by Udry et al. (2002) to harbor
a companion with a minimum mass of 17.5 M$_{\rm J}$. In Section 4 we compare our
findings both to similar spectroscopic studies and recently recalibrated
photometric estimates, and examine several subsamples of SWPs and field
stars with well-determined stellar parameters for potential trends in
metallicity, Galactic orbits, and stellar mass. We also offer interpretations
of these results in light of current theories of planetary system evolution.
We briefly summarize the main conclusions and discuss the implications of
our results in Section 5.

\section{Sample and Observations}
\label{obs}

High-resolution, high S/N ratio spectra of 24 stars were obtained with the 
2dcoude echelle spectrograph at the McDonald Observatory 2.7 m telescope 
using the same instrumental setup as described in Paper V.  An additional 
set of 7 southern stars were observed over two nights using the echelle spectrograph
of the CTIO 4 m Blanco telescope. Of these 31 targets, 25 were selected because they
had been reported to harbor substellar companions (see Table 1 for details of
the discovery papers), but had not been observed and analyzed by our group using
the methods we have consistently employed on previously announced SWPs. These
stars represented all of the known SWPs which were appropriate targets for our
analysis, but which we had not studied as of the dates of observations.
HD\,202206 was also included in this set, although its companion's 
minimum-mass estimate of 17.5 M$_{\rm J}$ suggests that it is more likely a 
brown dwarf than a planet. Two stars, HD\,95128 and HD\,117176,
which had been previously analyzed in Paper II using spectra from the McDonald 
Observatory 2.1 m telescope and the Sandiford cassegrain echelle,
were observed again using the 2.7 m instrument combination.
These repeat observations were made due both to the higher
quality of spectra which could be obtained, as well as to increase the self-consistency
of our SWP sample. We also obtained new spectra of HD\,16141, HD\,37124,
and HD\ 46375, each of which we had previously analyzed in Paper VI using Keck spectra.
The new spectra of these stars provides considerably expanded wavelength 
coverage, and allows us to employ our full atomic linelist in deriving results.
Several spectra of hot stars with high $v \sin i$ values were also obtained on
each night in order to divide out telluric lines in the McDonald and CTIO spectra.
Table 1 presents additional details of the SWPs observations.

We employed the same data reduction methods for echelle images as those 
detailed in Paper V to produce one-dimensional spectra covering most of 
the visible spectrum. Equivalent widths (EWs) were measured for the same set 
of Fe I and Fe II lines presented in Paper VI; the results are presented in 
Tables 2 and 3.\footnote{These tables are available in the electronic 
version of this paper.} EWs were also measured for several additional 
elements, and those results will be reported in a subsequent paper.

\section{Analysis}
\subsection{Spectroscopic Analysis}

Our method of determining the stellar parameters $T_{\rm eff}$, $\log g$, 
$\xi_{\rm t}$, and [Fe/H] is the same as that employed in our previous studies 
in this series, to which the reader is referred for more details. The results 
for our 31 program stars are given in Table 4.

Since we have not previously used the CTIO 4 m telescope for spectroscopic 
studies of SWPs, we need an independent check on the zero point of the 
derived [Fe/H] values. A comparison of our CTIO-based metallicity results
with the spectroscopic studies of Santos et al. (2000, 2001a) shows a mean
offset of $\Delta$[Fe/H]$=-0.01$, which is the same result we find in comparing
our entire McDonald-based SWPs sample to theirs. We therefore assume that 
the McDonald 2.7 m and CTIO 4 m spectra share a common zero point, and add no 
corrective offset to the CTIO-based results.

\subsection{Derived Parameters}

In previous papers in this series we have primarily employed the stellar 
evolutionary isochrones of Schaller et al. (1992) and Schaerer et al. (1993), 
along with the {\it Hipparcos} parallaxes (ESA 1997) and our spectroscopic 
$T_{\rm eff}$ estimates to determine masses, ages, and theoretical $\log g$ 
values.\footnote{Note, the theoretical $\log g$ values are derived from 
theoretical stellar evolutionary isochrones at the age which agrees with the 
observed T$_{\rm eff}$, M$_{\rm v}$, and [Fe/H] values.} In this work we have 
chosen to alter this method slightly by using the more recent Padova 
isochrone set (cf. Girardi et al. 2000, Salasnich et al. 2000). This was done 
in order to eliminate the offsets in $T_{\rm eff}$ and M$_{\rm v}$ 
that we had found necessary to reproduce solar values in the Schaller and 
Schaerer isochrone set (see Paper II), and also in order to take advantage of 
the greater range of age, mass, and metallicity values covered by the 
Padova isochrone set.

As before, we used our spectroscopic $T_{\rm eff}$ and {\it Hipparcos} 
M$_{\rm v}$ values to determine a best fitting isochrone for [Fe/H] 
values above and below the spectroscopic [Fe/H] of the star in question.
We then interpolated between these high- and low-metallicity isochrones to
estimate values for each star's age, mass, and theoretical $\log g$. In addition
to performing this updated procedure on the 31 new spectroscopic results presented
here, we re-analyzed all of the stars previously reported on in our SWPs series.
This set of homogeneously derived evolutionary parameters for 58 SWPs is presented
in Table 5, along with estimates of space velocities from {\it Hipparcos} data,
and additional age estimates based on the stellar activity index, $R^{'}_{\rm HK}$.

Several of the stars in Table 5 deserve note. Due to its location in the H-R
diagram, HD\,8574 presents two close but unique and equally viable solutions
for its derived parameters, at [Mass (M$_{\odot}$), Age (Gyr), $\log g$]
values of [1.16, 4.4, 4.20] or [1.11, 5.6, 4.18]. This degeneracy arises
due to the overlap in this region of the H-R diagram of the Main Sequence
and the post-main sequence blue ``hook'' of somewhat more massive stars.

An additional four stars - HD\,6434, HD\,37124, HD\,46375, and HD\,168746 - lie in
regions of the H-R diagram which are too red and/or too luminous to be well fit
by isochrones with appropriate metallicities and cosmologically reasonable ages.
Figure 1 presents H-R diagrams of these four stars along with Padova isochrones
of 14.1, 15.8 and 17.8 Gyr, selected to match as closely as possible our spectroscopic
[Fe/H] values for each star. Note that HD\,6434 and HD\,168746 are significantly
more metal-poor than stars which would lie precisely on the isochrones presented --
isochrones of the same ages but with proper, lower metallicities would have higher
temperatures and lie further to the left on these plots.

In contrast to the extreme ages predicted by the Padova isochrones, we note that
while HD\,6434 has a space velocity suggestive of old age, the other three stars
evidence relatively modest space motions, and furthermore three of the four possess
$R^{'}_{\rm HK}$ values which indicate ages less than 5 Gyr (HD\,168746 has no
published $R^{'}_{\rm HK}$ values). HD\,37124 and HD\,46375 have been previously
identified as possible `red stragglers' in Paper VI, and the very similar $T_{\rm eff}$
values we find in this work reinforce this claim. We have previously proposed that
these stars might be overluminous due to the presence of unseen stellar companions;
however, a recent adaptive optics search for nearby stellar companions to SWPs by
Luhman and Jayawardhana (2002) failed to detect any accompanying objects capable of
producing the necessary luminosity enhancements for HD\,37124 or HD\,46375.
In light of these unusual and conflicting findings, we recommend that HD\,6434 and
HD\,168746 also be examined for close stellar companions, and that all four of these
`red stragglers' be targeted for follow-up studies to determine the source of their
unorthodox characteristics.

\section{Discussion}
\subsection{Comparisons with Other Studies}

Continuing efforts to characterize F, G, and K dwarfs in the solar neighborhood,
as well as the considerable attention focused on the discovery and subsequent
follow-up studies of SWPs themselves, have resulted in the development of multiple
independently produced sets of spectroscopic and photometric estimates of basic
stellar parameters for these objects. In the following subsections we compare our
results to those of several recent studies.

\subsubsection{Metallicity}

Spectroscopic values of [Fe/H] have been reported by Santos et al. (2000, 2001a, 2002)
for most of the stars in the present study. In comparing [Fe/H] values between
the two sets, we find a mean difference of [Fe/H]$_{\rm Here} - $[Fe/H]$_{\rm Santos}
=-0.009 \pm 0.007$, with a standard deviation of $0.039$. This very close agreement
is not a complete surprise, given the considerable similarity in the analysis method
used by our group and that of Santos et al., but is nevertheless an encouraging indicator
of the consistency and precision of our results. Additional comparisons of our spectroscopic
[Fe/H] values for SWPs with those reported by several other investigators can be found in
Gonzalez (2003).

Martell \& Laughlin (2002) and Kotovena et al. (2002) offer newly calibrated
equations for estimating metallicities from photometric data. The former utilizes
Str\"omgren indices to estimate [Fe/H], while the latter employ luminosities (M$_{\rm v}$) 
andJohnson-Cousins [$B-V$] colors to determine [Fe/H] and is specifically formulated
for later spectral types (5.5 $<$ M$_{\rm v}$ $<$ 7.3). We used these to calculate
photometric [Fe/H] values for a set of 69 stars with substellar
radial velocity (RV) companions for which previous spectroscopic metallicity results
exist, using the Kotoneva method for all stars with M$_{\rm v}$ $>$ 5.50, and the
Martell/Laughlin equation for all brighter stars. We find a mean offset of 
[Fe/H]$_{\rm Spec} - $[Fe/H]$_{\rm Phot} = 0.10 \pm 0.01$, with 
a standard deviation of $0.10$ between the spectroscopic and photometric estimates.
Figure 2 shows the relation of this offset with $T_{\rm eff}$, 
$\log g$, and [Fe/H], which reveals no significant trend with any of these 
parameters. This suggests a zero-point shift in [Fe/H] between the calibrating 
spectra of these photometric studies and the spectroscopic studies discussed here.

\subsubsection{Surface Gravity and Mass}

For an independent check on our $\log g$ and mass estimates, we turn to
Allende Prieto \& Lambert (1999, hereafter APL99), who utilized the isochrones
of Bertelli et al. (1994) and {\it Hipparcos} data to generate estimates of mass and
$\log g$ for just over 17,000 stars within 100 pc of the Sun. Of the 58 SWPs listed in
Table 5, 53 are also included in the APL99 sample, allowing us to calculate tables of
deviations between our two data sets and examine these differences for trends.
We find that both our spectroscopic and theoretical estimates of $\log g$ compare
favorably with those of APL99, with the following mean offsets: $\log g_{\rm APL99}
- \log g_{\rm spec} = -0.01 \pm 0.02$; $\log g_{\rm APL99} - \log g_{\rm
theory} = 0.03 \pm 0.01$.

Our mass estimates are also in very good agreement with APL99, with a mean
difference of M$_{\rm APL99} - $M$_{\rm spec}= \Delta M = -0.04 \pm 0.02$ M$_{\odot}$.
However, we do find a relationship between the deviations in mass estimates and
[Fe/H], which we attribute to the exclusive use of solar metallicity isochrones by APL99.
We have derived the following second-order equation to correct for this:
\begin{equation}
\Delta M = 0.2234 [Fe/H]^2 - 0.3023 [Fe/H] - 0.0095
\end{equation}
Provided [Fe/H] values from other sources are available (e.g. from Str\"omgren colors, 
as in Martell \& Laughlin (2002)), this correction may be applied to the 
APL99 data to provide somewhat more accurate estimates of stellar mass.

\subsubsection{Effective Temperature}

A comparison of our values for $T_{\rm eff}$ with those of APL99 show a mean
difference of $T_{\rm eff (spec)} - T_{\rm eff (phot)} = -34 \pm 16$ K, with a
standard error of 120 K, a reasonable match given the disparate methods employed
and the lower precision of the temperatures reported in that work.
Similarly, spectroscopic estimates of $T_{\rm eff}$ available from
the Geneva Extrasolar Planet Search Programme website (Santos et al.
(2001a), Santos et al. (2002), Perrier et al. (2002)) for 30 of the 31 stars
analyzed here are also in close agreement, with a mean difference of $-40 \pm 10$ K
and a standard error of 55 K. For further comparisons of our temperature estimates
with those of additional spectroscopic studies see Gonzalez (2003).

\subsubsection{Age}

Independent age estimates based on $R^{'}_{\rm HK}$ values are available for 44 of
the 58 stars listed in Table 5 (the remainder either lack published $R^{'}_{\rm HK}$
values or possess highly uncertain spectroscopic ages). Figure 3 shows the deviations
between our Padova ages and those based on $R^{'}_{\rm HK}$ estimates as a function of
$T_{\rm eff}$, and we note a distinct trend of relatively lower $R^{'}_{\rm HK}$ age
with increasing stellar temperature, in agreement with a systematic offset in
$R^{'}_{\rm HK}$ age with spectral type previously noted amongst SWPs by Barnes (2001).
Although the $\log R^{'}_{\rm HK}$ values quoted in Table 5 may not be representative of
the true activity levels of the star in question (as such reliability requires observations
over several years to account for variability in stellar activity), it is unclear how such
observational uncertainties could lead to the trends seen here and amongst the SWPs studied
in Barnes (2001). A likely source of systematic offset between the two age estimates may
be found in the different isochrone sets used; the $R^{'}_{\rm HK}$ age calibration we
employ is from Donahue (1993), which follows the method of Soderblom et al. (1991) who
utilized the isochrones of Maeder (1976). In light of these points and those raised by
many other studies (c.f. Donahue (1998)), the evidence we present strongly suggests a
need to revise the $R^{'}_{\rm HK}$ age calibration.

\subsection{Spectroscopic Sample}

Potential correlations between the metallicities and other physical characteristics
of SWPs have been remarked upon since shortly after the discoveries of these systems
(c.f. Paper I, Marcy and Butler (1997)), and more recent studies have pursued these
possibilities as the sample of SWPs has grown (Paper VI, Reid (2002), Santos et al.
(2002)). We continue this by examining several of these trends below in light of the new
data presented, using a ``Spectroscopic Sample'' of SWPs constructed in the manner of and
for the reasons described in Paper VI. Accordingly, it includes only SWPs for which we
have derived spectroscopic [Fe/H] values, along with 19 SWPs with parameters from Santos et
al. (2002) (owing to the similarity of our methods and results -- see above). We further
exclude all stars whose companions have a minimum mass greater than 11 M$_{\rm J}$, as well
as seven biased objects, specifically added to Doppler search programs due to their high
metallicities: BD -10 3166, HD\,2039, HD\,4203, HD\,30177, HD\,73526, HD\,76700, HD\,108874.
Lastly, we have chosen to continue to include in our Spectroscopic Sample
HD\,192263, whose planetary companion has been challenged by Henry et al. (2002). While
we believe the evidence in that work to be compelling, the discovery team of HD\,192263
has not yet retracted its findings \footnote{Although the same group {\it has} recently
retracted claims of RV companions to HD\,13507 and HD\,223084 (neither of which have been
included in any of our statistical studies).}; we will, however, make careful note of the
presence of HD\,192263 in any figures or derived statistical quantities. In sum, these
constraints yield a Spectroscopic Sample which contains a total of 71 SWPs.

\subsubsection{Metallicity Distribution}

As first suggested in Paper I, and later confirmed by a number of
detailed studies (c.f. Paper VI, Santos et al. (2001a), Reid (2002)), the 
current population of stars hosting radial velocity companions is metal-rich by 
$\sim0.25$ dex in comparison to similarly selected samples of field stars not known
to harbor planetary or brown-dwarf companions (exact values of this offset vary slightly
depending on the nature of the SWPs and control samples). The data presented
here continue to support this now well-established feature of SWPs, with a mean
metallicity of [Fe/H]$=+0.12 \pm 0.02$ for our Spectroscopic Sample, compared with
an average [Fe/H] of $-0.10 \pm 0.03$ for the stars-without-planets sample of Santos
et al. (2001a). This latter sample shares the same zero point as the larger CORALIE
sample presented in Figure 3c of that work, and we compare the distribution of
[Fe/H] among our Spectroscopic Sample with the $\sim 1000$ CORALIE stars in Figure 
4.
 
We do note, however, that while the mean [Fe/H] for SWPs is still significantly higher
than that of the general population of stars in the solar neighborhood, it is 
slightly lower than that of the Spectroscopic Sample studied in Paper VI. This raises the
possibility that as the nature of the planets detected by Doppler searches evolves to
include both less massive planets and systems with longer orbital periods, the metallicity
distribution of SWPs may change as well. We illustrate this in Figure 5, which overplots
histograms  of metallicities from the 38 stars in the Spectroscopic Sample of Paper VI with
those of the 25 SWPs reported here (the updated [Fe/H] values for HD\,16141, HD\,37124, 
HD\,46375, HD\,95128, and HD\,117176 are incorporated into the Paper VI 
histogram data). This may suggest a more subtle relationship between host star 
metallicity and the periods and/or masses of planetary companions -- a 
possibility we will investigate further below -- and is a development which should
be monitored closely.

\subsubsection{Young SWPs}

In keeping with our analysis in Paper VI, we have prepared a subsample of SWPs which,
in addition to meeting the criteria for inclusion in the Spectroscopic Sample, have
spectroscopic age estimates of less than 2.0 Gyr. This sample is shown in Figure 6, where
again in analogy with Paper VI we plot [Fe/H] against mean Galactocentric distance, $R_{\rm M}$,
relative to the position of the Sun, $R_{\rm 0}$, for both young SWPs and a sample of
young field stars from Edvardsson et al. (1993) (HD\,192263, which is young, but whose status
as a SWP is in question, is plotted separately as a square symbol). The dashed line is a 
least-squares fit to the sample of field stars, and we find that with the 
exception of HD\,130322, all of the SWPs lie above this trend.

In a recent study of the Galactic orbits of SWPs Barbieri \& Gratton (2002) find a similar
result, in that at any Galactic radius SWPs possess metallicities above the mean of
nearby field stars. The authors suggest that this is best explained by a scenario in which
the presence of companions is responsible for the observed high metallicities. Although comparisons
between our studies are not exact -- our young SWPs sample does not have as large a range in
[Fe/H] values as the age-independent Barbieri \& Gratton study, but suffers less from the effects
of orbital diffusion -- we do confirm this observational result. However, in contrast to
Barbieri \& Gratton we also find evidence of a difference in the slope of the
metallicity-Galactocentric radius relation between SWPs and the field star
sample. Because SWPs are so much more common among metal-rich stars, the trend with
Galactocentric radius is observed to be steeper than that amongst the field population.
This could be interpreted as supporting a hypothesis in which the high metallicity
is reponsible for the presence of companions.

\subsubsection{Planetary Characteristics}

Figures 7 and 8 present (respectively) the eccentricities and $M {\sin i}$ values
of the planets in the Spectroscopic Sample, shown in each case
as a function of host star metallicity (for SWPs with more than one detected companion,
the eccentricity of the innermost planet and the combined $M {\sin i}$ value of all
planets in the system is adopted). In general agreement with Santos et al. (2002), no
significant trend is visible for either distribution, although the upper envelope of
each is observed to increase somewhat with higher values of host star [Fe/H].

\subsection{The California \& Carnegie Planet Search Sample}

As discussed at length in Reid (2002), Santos et al. (2001a, 2002),
and Gonzalez (2003), investigations of possible uniquely characteristic observational
features of SWPs face the difficulty of the lack of a comprehensive, volume-limited
control sample. We can mitigate this somewhat by strongly
limiting the scope of our query, as in the case of the young SWPs; however, in order
to evaluate more global trends among SWPs, we require a more complete understanding
of the stars being searched for RV companions. Figure 4 approaches this goal with its
comparison of the metallicities of SWPs to the CORALIE sample described in Santos et al. 
(2001a), but no additional data beyond the relative frequency of stars as 
a function of metallicity has been published to date on that control sample, 
rendering further inquiries into other physical characteristics difficult, if 
not impossible.

Mass is one important property of SWPs not yet examined in detail. In order to determine
any potential trend in the masses of SWPs, we prepared a control sample from the list of
889 stars currently under observation by the California and Carnegie Planet Search Program
(CCPSP), for which accurate mean RV data has recently been published by Nidever et al.
(2002). This set of stars was selected because its detailed publication allows for further
investigation of its member stars by additional methods, and its inherent inclusion of
those SWPs found by the CCPSP enables ready comparison of those stars discovered to harbor
RV companions with similarly studied stars which do not yet show such evidence.

From the initial data set of 889 stars we removed all stars with visual magnitudes
fainter than 8.0, the current stated limiting magnitude of the CCPSP, as well as all stars
whose {\it Hipparcos} parallax estimates possessed errors greater than 20\% of their
nominal value (this latter restriction was made in order to insure the reliability of
subsequent estimates of M$_{\rm v}$). Additionally, stars with M$_{\rm v} < 2.0$ and
$B-V > 0.90$ were omitted, as were stars with M$_{\rm v} > 7.73$, in order to restrict
the comparison sample to an F, G, and K dwarf subset. This process excluded several
subgiant stars from the CCPSP Subsample, but no members of the SWPs cohort. HD\,4203,
HD\,49674, and HD\,108874 were also removed as they were added to the CCPSP solely due
to their high values of [Fe/H]. Lastly, we removed an additional 9 stars for which no additional
published spectroscopic or photometric data were available to provide estimates of [Fe/H].
In total, this sample (hereafter, the ``CCPSP Subsample'') contains 496 stars, 36 of which
have been reported by the CCPSP to host RV companions with minimum masses less than
11 M$_{\rm J}$.

\subsubsection{Metallicities}

We calculated values of [Fe/H] for the CCPSP Subsample using the previously described
photometric recipes of Martell \& Laughlin (2002) and Kotoneva et al. (2002),
employing the Martell/Laughlin equations for all stars brighter than M$_{\rm v} = 5.5$,
and the Kotoneva method for those dimmer. In keeping with our findings of an 0.10 dex
offset between these photometric estimates and our spectroscopic results, we added
0.10 dex to the photometric [Fe/H] values of all stars in the CCPSP Subsample. Figure
9 presents a histogram of the relative contribution of SWPs to the entire CCPSP
Subsample as a function of metallicity. As we have seen in similar analyses by other
investigators (Santos et al. 2001a, Santos et al. 2002, Reid 2002), a significantly
higher fraction of the metal-rich CCPSP Subsample possess RV companions relative to the
more metal-poor stars (although the highest metallicity bin shown does suffer somewhat
from low number statistics, as there are only 3 stars in the CCPSP Subsample with such
high values of [Fe/H]).

\subsubsection{Stellar Masses}

In order to readily generate mass estimates for the CCPSP Subsample, we performed a multiple
linear regression analysis on the M$_{\rm v}$, $B-V$, and [Fe/H] data points contained within
a Padova isochrone set with a fixed age of 3.98 Gyr and [Fe/H] values ranging from $-0.38$
to $+0.32$ dex. This yielded the following equation relating stellar mass, M$_{\rm v}$,
$B-V$, and [Fe/H]:
\begin{equation}
M / M_{\odot} = 1.61437 - 0.153824 M_{\rm v} + 0.116028 [Fe/H] + 0.216553 (B-V)
\end{equation}
We employed equation (2) on the full CCPSP Subsample, and a comparison of the mass
values thus generated with those calculated for the same SWPs using the Padova isochrones
shows a mean difference of
M$_{\rm Spec. Est.} - $M$_{\rm Linear Est.}= -0.01 \pm 0.01$ M$_{\odot}$, with a
standard deviation of 0.09 M$_{\odot}$. The left panel of Figure 10 presents a histogram
distribution of the masses of the CCPSP subsample and the CCPSP SWPs, while the right panel
presents the same information as a cumulative frequency distribution. A Kolmogorov-Smirnov
analysis of the latter indicates only a $14\%$ probability that the two samples were drawn from
the same population, suggesting the presence of a mass bias in the detection of SWPs amongst
the full CCPSP Subsample.

There are many potential sources for such biases -- for example, given the relatively high mean
metallicity of SWPs and the tendency for more massive stars to be younger and
therefore more metal-rich. It is plausible that there should be a bias favoring 
the detection of SWPs around more massive stars. However, the orbital velocity of a star 
harboring a companion is an inverse function of that star's mass, so that for any given 
companion mass, a more massive host star will be accelerated less by their mutual gravitational
attraction, and thus will have a lower orbital velocity with an accordingly lower variability
in its measured doppler signal. More precisely, the velocity amplitude goes as 
$K_{\rm SWPs} \propto M_{\rm SWPs}^{-2/3}$ (Cumming et al. 1999). Additionally, more
massive stars have fewer absorption lines and are more likely to possess higher rotational
speeds, with both factors greatly reducing the probability of detecting the doppler signal
from a companion (Bouchy et al. 2001).

Figure 11 attempts to address several of these possible concerns. Each panel presents
histograms of the relative contribution of CCPSP SWPs to the full CCPSP Subsample as a
function of mass (solid line), as well as mean results from Monte-Carlo simulations (dashed
line) with $\sim 20,000$ trial selections of `pseudo-SWPs' samples, each equal in size to the
true CCPSP SWPs sample (36 stars). Assuming a Poisson distribution of error in bin membership
for the true SWPs, the formal uncertainty in each bin is $\leq 0.04$, while the standard
deviations of the Monte-Carlo distributions are $\sim 0.02$ for all but the extreme upper
and lower mass bins, which suffer from low numbers and have uncertainties of $\sim 0.10$.
Panel (a) presents both distributions with no attempts to correct for any biases; as expected,
each bin in the Monte-Carlo distribution contains a $\sim 7$ percent `pseudo-SWPs' component,
in accordance with the overall ratio of SWPs to the full CCPSP Subsample, and the bias
suggested by Figure 10 is again evident.

The remaining three panels of Figure 11 consider corrections for possible biases. For panel
(b), the true SWPs distribution is the same, but the Monte-Carlo selection process has
been altered to force the `pseudo-SWPs' samples to mimic precisely the metallicity distribution
observed amongst the true SWPs (i.e., the same number of stars in each 
metallicity bin). This eliminates from the simulated data those CCPSP stars
below 0.7 M$_{\odot}$, whose metallicities are lower than any of the observed CCPSP SWPs, but
also leads to a large overestimation of the contribution of the high mass stars as compared to
the true SWPs distribution. This latter result argues strongly against high [Fe/H] values
as the sole bias in the observed mass function.

Panels (c) and (d) attempt to address the previously mentioned biases against detecting SWPs
around higher mass stars. In each case, and for both the observed and simulated data, we 
compensated for the radial velocity bias by applying a multiplicative correction of
M$_{\rm Stellar}^{2/3}$ to the frequency of SWPs (or `psuedo-SWPs') in each mass bin,
resulting in an increase in the relative contribution of stars greater than 1.0 M$_{\odot}$
and a decrease in those below 1.0 M$_{\odot}$. We note that though this is theoretically
precise, it is probably a somewhat overly strong correction for this effect, as the 
doppler signals of the observed SWPs have relatively high S/N. However, as we
make no effort to compensate for the several additional biases presented against higher mass
stars (as they are difficult to directly and reliably quantify as functions of mass), we are
confident that this correction as applied offers a reasonable resolution. Panel (c) of Figure
11 presents the mass-bias corrected distributions of both the simulated and observed data,
while panel (d) considers both the mass-bias correction as well as the metallicity constraints
as applied in panel (b), and therefore represents our best estimate of the `true' mass function.

Evident immediately in all of the plots in Figure 11 is a steep dropoff in the percentage of
stars less than 1.0 M$_{\odot}$ harboring known RV companions -- only 6 SWPs have been detected
from the 148 CCPSP Subsample stars less massive than the sun, while 17 SWPs have been found
amongst the 152 stars in the CCPSP Subsample with masses between 1.0 and 1.1 M$_{\odot}$.
The sharp downturns below 0.8 M$_{\odot}$ and above 1.3 M$_{\odot}$ are also suggestive,
but as there are only 36 CCPSP stars in the mass bins at these extremes, even a high estimate
of a 10\% contribution by SWPs would predict only 1 or 2 SWPs at these masses. While we
must emphasize that these suggestive results are only marginally statistically significant at
this stage, if any of these trends are indicative of the true mass function of SWPs,
they provide very strong constraints on models of planetary system formation and evolution, and
impact profoundly the possible distribution of SWPs in the Galaxy.

\subsection{Sources of Trends}

Gonzalez (2003) summarizes the three currently debated mechanisms for explaining
the high [Fe/H] seen among SWPs -- the {\it primordial}, {\it self-enrichment}, and
{\it migration} hypotheses. The trends seen above in the Galactic orbits of SWPs
would seem to support the findings of Barbieri \& Gratton (2002) favoring the
self-enrichment scenario, but the change in the slope of the metallicity-galactocentric
distance relationship does suggest some dependence upon the value of [Fe/H], a prediction
of the primordial hypothesis. The latter also finds support in the high metallicity of
the subgiants HD\,27442, HD\,38529, and HD\,177830, whose outer convection 
zones should be deep enough to effectively dilute any reasonable amount of 
self-enrichment. However, the addition here of several SWPs to the metal-poor 
subset with [Fe/H] $< -0.20$ continues to provide a lower limit to any 
critical metallicity necessary for planet formation.

Figure 12 presents an analysis of our theoretical and spectroscopic $\log g$ values. 
This offers a possible insight into the self-enrichment scenario, given that a star 
with an enriched outer convection layer will have a more metal-poor interior 
than estimated spectroscopically. This results in a lower mass (and therefore 
$\log g$) estimate than that given by evolutionary models for a star whose 
outer atmosphere shares the metal abundance of its interior (Ford et al. 
1999). While this effect is small and difficult to reliably detect at
the current level of resolution\footnote {An 0.2 dex enhancement in [Fe/H] in the convection
layer of 51\,Peg, for example, would yield only an 0.05 dex offset in $\log g$.}, we note
that our spectroscopic $\log g$ estimates are found to be on average higher than our 
theoretical$\log g$ estimates, in contradiction to this simple model.

The possible trends with SWPs mass seen in Figures 10 and 11 suggest the hypothesis that
large, relatively close-orbiting planets may be preferentially found around more massive
stars. However, some of this observed distribution of SWPs with stellar mass may be a
reflection of the increased probability of finding SWPs orbiting stars with higher [Fe/H],
as higher mass stars are on average younger, and therefore generally more metal-rich than
lower mass stars. We explore this degeneracy in Figure 13, which presents [Fe/H] as a
function of mass for both the full CCPSP Subsample and the subset of CCPSP
SWPs. We note that the {\it most} metal-rich SWPs are found just
above 1.0 M$_{\odot}$, not amongst the SWPs with the highest masses, and that 
there are large numbers of metal-rich, higher mass stars not in the SWPs
subset. These points qualitatively argue against metallicity being the sole 
cause for the observed mass trend; more thorough statistical examinations 
of the data set as it grows should be performed to properly determine the level to which these
two stellar properties bias each other.

If there is in fact a mass dependence on the likelihood of possessing a planetary
companion, there are many potential avenues of inquiry to be explored -- are massive stars
perhaps more likely to suffer the effects of planetary migration via interactions with a more
massive disk, or perhaps to directly form through disk instabilities giant planets in $< 5$
AU orbits? Or are they simply more likely to form more massive, more easily detected planets,
or perhaps more planets in general? Answers to these questions will most likely be found via
continued searches for planets around stars of various masses, but with similar ages,
metallicities, and environmental histories, such as the study of the Hyades by Cochran et al.
(2002). The rapidly developing field of modelling planetary system formation may also yield
insight into the underlying physical causes of the observed mass trend. Given its potential
to powerfully constrain theories of planetary system evolution as well as the overall
distribution of SWPs in our Galaxy, we strongly urge research groups to follow up on this
preliminary and potentially critical finding.

\section{Conclusions and Implications}

In this paper we have presented new spectroscopic analyses of 31 SWPs, including 25 SWPs
which had not been previously examined in the context of our ``Stars-with-Planets'' series
of papers. Where overlap exists with other, similar research, we find very good agreement
between our results on the basic stellar parameters of $T_{\rm eff}$, $\log g$, and [Fe/H].
We further utilized our spectroscopic data and the Padova isochrone set to derive ages, masses,
and theoretical $\log g$ values for all of the SWPs examined in our series of papers. These
data revealed a number of very interesting and unusual stars, including four ``red stragglers'':
HD\,6434, HD\,37124, HD\,46375, and HD\,168746. HD\,46375 is particularly unusual, in that it
shows evidence of being both very old and very metal-rich. We recommend that these stars be
the targets of detailed studies to determine the source(s) of their abberant physical
characteristics.

When we combine the results here with that of our previous work and those of other very similar
spectroscopic studies, we find a preponderance of evidence supporting the hypothesis that SWPs
are preferentially metal-rich compared to control samples of field stars, as first proposed
in Paper I, and additionally confirmed by numerous research efforts (Paper VI, Santos et
al. 2001a, Santos et al. 2002, Reid 2002). This observed enhancement in [Fe/H] over similar
nearby stars is seen at all Galactic radii amongst a subsample of young SWPs, a prediction of
the {\it self-enrichment} model, but also appears to show some dependence on base 
metallicity, a feature of the {\it primordial} model (and arguably the {\it migration} model as 
well). No strong evidence was found linking stellar [Fe/H] to planetary characteristics,
however, and a proposed test of the self-enrichment model via discrepancies in determinations
of $\log g$ measured through direct spectroscopy and by application of stellar evolutionary
models were found to be largely inconclusive. While it is almost certainly the case that there
is some minimum primordial metallicity necessary for the formation of a planetary system, and
that once formed any planetary system will almost certainly enrich its host star with metals,
the degree to which either of these is responsible for the observed metal-rich nature of SWPs
remains in question. Pinsonneault et al. (2001) and Santos et al. (2002) provide compelling
evidence against widespread self-pollution as the source of the high mean metallicity of SWPs,
while Gratton et al. (2001), Murray \& Chaboyer (2002), Smith et al. (2002), and Laws \& Gonzalez
(2001) offer similarly significant evidence supporting non-negligible levels of self-pollution.

Finally, we used a multiple linear regression method and data from Padova isochrones
to estimate masses for 496 stars currently targeted for doppler studies by the California
\& Carnegie Planet Search Program, employing slight recalibrations to the recent photometric
studies of Martell \& Laughlin (2002) and Kotoneva et al. (2002) to derive necessary [Fe/H]
values. A comparison of this sample with the subset of known SWPs within this group demonstrated
both the (expected) trend towards more frequent occurrence of SWPs amongst high metallicity
stars, as well as a somewhat increased probability of detection of RV companions in orbit
around more massive stars. While this last finding is highly tentative and marginal in its
statistical significance, its ramifications are broad and deep. If our results are in fact
indicative of the underlying mass function of SWPs, then this strongly constrains models of
planetary system formation to prescriptions which preferentially populate stars 
of higher mass with planets typical of those being discovered by current 
doppler surveys. Whether the deficit of giant planets around low mass stars 
is the result of a real physical effect or an artifact of the metallicity 
trend, the implications for the total number of SWPs in the Galaxy are similar. 
Given the fact that the overwhelming majority of stars in our Galaxy are less 
massive than the sun, any tendency for such stars to be without large planetary 
companions has a profound impact on the total number of planets in our Galaxy. 
Were it to be the case, as suggested in Wetherill (1994) that such large worlds 
are necessary for the development of intelligent life in a planetary system,
then a lack of these giants around the most common stars in most galaxies may 
have important implications on estimates of the number of possible harbors of 
civilizations in the universe.

\acknowledgements
  
We would like to thank Kevin Covey for his work on software that greatly aided
this analysis, and Tammy Ann Baker Laws for assistance in the preparation of tables.
This research has made use of the Simbad database, operated at CDS, Strasbourg, France,
as well as Jean Schneider's, Geoff Marcy's, and the Geneva Research Group's extrasolar
planets web pages. This work has been supported by a the University of Washington
Astrobiology Program and a grant from the National Astrobiology 
Institute. Additionaly, Sudhi Tyagi was supported by the Space Grant Program 
at the University of Washington, and Kyle M. Walker was supported by a 
Research Experience for Undergraduates grant from Iowa State University.

\clearpage

\section{FIGURE CAPTIONS}

\figcaption{H-R diagrams of 4 SWPs with ``red straggler'' characteristics are shown.
Crosshairs in each plot are centered on values for M$_{\rm v}$ and 
$\log T_{\rm eff}$ given in the text, with dimensions equal to quoted
$1-{\sigma}$ uncertainties. Padova isochrones for ages of 14.1 Gyr (solid line),
15.8 Gyr (dashed line), and 17.8 Gyr (dot-dashed line)
are overplotted for the following metallicities: HD\,6434, $-0.38$; HD\,37124,
$-0.38$; HD\,46375, $+0.32$; HD\,168746, $+0.00$.} 

\figcaption{Differences between Spectroscopic and Photometric estimates of [Fe/H], 
shown as functions of $T_{\rm eff}$ (upper panel), $\log g$ (middle panel), and
Spectroscopic [Fe/H] (lower panel).}

\figcaption{Differences between SWPs ages derived using the Padova isochrone set
and ages calculated with the $R^{'}_{\rm HK}$ calibration, shown as a function of
$T_{\rm eff}$.}

\figcaption{The relative frequency of SWPs (dashed line) and field stars (solid
line) amongst the at varying metallicities. The SWPs are from the Spectroscopic
Sample and consist of 71 stars hosting radial velocity companions with spectroscopic
[Fe/H] values measured by our group or by the Geneva group (see text for references),
and companion masses below 11 M$_{\rm J}$. Seven SWPs, listed in the text, are omitted
from this comparison, as they were specifically added to radial-velocity programs
because of their suspected high metallicity. The questionable SWP HD\ 192263 is
included, and contributes 1.5\% to the $-0.10 < $[Fe/H]$ < 0.00$ bin. The field star
sample consists of $\sim$ 1000 stars with metallicities determined by CORALIE (see
fig. 3c of Santos et al. 2001a).}

\figcaption{A comparison of the metallicity distribution of the SWPs samples
presented in Paper VI (solid line) and here (dashed line). Five stars reported
on in Paper VI (see Table 4) and re-analyzed in the current work are shown as
members of the Paper VI sample, but with their updated [Fe/H] values.}

\figcaption{[Fe/H] vs. mean Galactocentric distance, $R_{\rm M}$, relative to the
position of the Sun, $R_{\rm 0}$, for SWPs (crosses) with Padova age estimates less
than 2.0 Gyr and a sample of field stars from Edvardsson et al. (1993) meeting the
same age criterion (dots). HD\,192263, whose SWP status is in question, is 
plotted separately as a square symbol here and in the following two figures. 
The dashed line is a least-squares fit to the sample of field stars.}

\figcaption{Eccentricity of planetary orbits versus host star [Fe/H] for the 
71 SWPs in the Spectroscopic Sample (for SWPs with more than one detected 
companion, the eccentricity of the innermost planet in the system is adopted).}

\figcaption{Mass of planetary companions ($M {\sin i}$) verus host star [Fe/H] 
for the 71 SWPs in the Spectroscopic Sample (for SWPs with more than one 
detected companion, the combined $M {\sin i}$ value for all planets in the 
system is adopted).}

\figcaption{The relative contribution as a function of metallicity of the CCPSP
subsample of 36 SWPs to the full CCPSP Subsample of 496 stars (all drawn from the larger
CCPSP target list of Nidever et al. (2002)). Formal uncertainties assuming a Poisson
distribution of error in binning are $\pm 0.04\%$ in each bin. The exceptional value
centered on [Fe/H]$=+0.55$ is the result of one SWP discovered amongst the only three
CCPSP stars with such high metallicities.}

\figcaption{The distribution of masses amongst the entire CCPSP Subsample (dashed line)
and the subset of SWPs contained therein (solid line), shown in histogram form in the
left panel and as a cumulative frequency plot in the right panel. A Kolmogorov-Smirnov
test of the two distributions indicate a $14\%$ probability that they are drawn
from the same population.}

\figcaption{The relative contribution as a function of mass of the 36 SWPs 
in the CCPSP Subsample compared to the full CCPSP Subsample of 496 stars (solid line).
Overplotted are results from Monte Carlo simulations of the mass distributions (dashed
line, see text for details). Panel (a) -- raw distribution of observed and simulated
SWPs mass data; Panel (b) -- raw distribution of observed SWPs and metallicity constrained
simulated data; Panel (c) -- mass-bias corrected distribution for both observed and simulated
SWPs; Panel (d) -- same as panel (c), but also including the same metallicity constraints
for the simulated data, as applied in panel (b). Formal uncertainties in the observed SWPs
mass data, assuming a Poisson distribution of error in binning, are $\pm 0.04$ in each bin.
Standard deviations of each bin in the simulated data are $\pm 0.02$ in all but the extreme
upper and lower mass bins, which have uncertainties of $\sim 0.10$ owing to low number
statistics.}

\figcaption{Deviations between theoretical and spectroscopic estimates of $\log g$ as
a function of spectroscopic $\log g$ for the 58 SWPs in Table 5.}

\figcaption{Photometric [Fe/H] as a function of stellar mass for 496 stars in the CCPSP
Subsample (crosses) and the 36 memers of the SWPs subset (filled circles). Dashed lines
represent approximate selection boundaries, following the M$_{\rm v}$ and color cutoffs of the
CCPSP Subsample described in section 4.3.}

\clearpage

\begin{deluxetable}{lccccccc}
\tablecaption{Observing Log}
\scriptsize
\setlength{\tabcolsep}{0.03in}
\tablewidth{0pt}
\tablehead{
\colhead{Star} & \colhead{UT Date} & \colhead{Observatory} & 
\colhead{Wavelength} & \colhead{Resolving} & 
\colhead{S/N\tablenotemark{a}} &
\colhead{Observer} & \colhead{Discovery}\\
\colhead{(HD)} & \colhead{(mm/dd/yy)} & \colhead{} & 
\colhead{Range (\AA)} & \colhead{Power} & \colhead{(per pixel)} &
\colhead{code\tablenotemark{b}} & 
\colhead{code\tablenotemark{c}}
}
\startdata
82943 & 12/14/2000 & McD 2.7-m & 3700-10000 & 63,000 &375 & CL &1 \nl
33636 & 12/14/2000 & McD 2.7-m & 3700-10000 & 63,000 &340 & CL &1, 2 \nl
22049 & 12/14/2000 & McD 2.7-m & 3700-10000 & 63,000 &240 & CL &3 \nl
4208 & 12/15/2000 & McD 2.7-m & 3700-10000 & 62,000 &325  & CL &2 \nl
190228 & 12/15/2000 & McD 2.7-m & 3700-10000 & 62,000 &360  & CL &4 \nl
1951019A & 12/15/2000 & McD 2.7-m & 3700-10000 & 62,000 &300  & CL &5 \nl
16141 & 12/15/2000 & McD 2.7-m & 3700-10000 & 62,000 &320  & CL &6 \nl
46375 & 12/15/2000 & McD 2.7-m & 3700-10000 & 62,000 &350  & CL &6 \nl
37124 & 12/15/2000 & McD 2.7-m & 3700-10000 & 62,000 &300  & CL &7 \nl
213240 & 04/06/2001 & CTIO 4-m & 5850-8950 & 35,000 &320 & NS &8 \nl
108147 & 04/06/2001 & CTIO 4-m & 5850-8950 & 35,000 &260 & NS &9 \nl
121504 & 04/06/2001 & CTIO 4-m & 5850-8950 & 35,000 &265 & NS &1 \nl
160691 & 04/05/2001 & CTIO 4-m & 5850-8950 & 35,000 &230 & NS &10 \nl
169830 & 04/05/2001 & CTIO 4-m & 5850-8950 & 35,000 &225 & NS  & 11\nl
83443 & 04/06/2001 & CTIO 4-m & 5850-8950 & 35,000 &280 & NS &12 \nl
27442 & 04/06/2001 & CTIO 4-m & 5850-8950 & 35,000 &195 & NS &13 \nl
4203 & 12/02/2001 & McD 2.7-m & 3700-10000 & 60,000 &250 & CL,GG  &2 \nl
8574 & 12/02/2001 & McD 2.7-m & 3700-10000 & 60,000 &405 & CL,GG  &1 \nl
19994 & 12/02/2001 & McD 2.7-m & 3700-10000 & 60,000 &620 & CL,GG  &1 \nl
28185 & 12/02/2001 & McD 2.7-m & 3700-10000 & 60,000 &245 & CL,GG  &8 \nl
50554 & 12/02/2001 & McD 2.7-m & 3700-10000 & 60,000 &390 & CL,GG  &1 \nl
68988 & 12/02/2001 & McD 2.7-m & 3700-10000 & 60,000 &250 & CL,GG  &2 \nl
95128 & 12/02/2001 & McD 2.7-m & 3700-10000 & 60,000 &550 & CL,GG  &14 \nl
117176 & 12/02/2001 & McD 2.7-m & 3700-10000 & 60,000 &335 & CL,GG  &15 \nl
106252 & 12/02/2001 & McD 2.7-m & 3700-10000 & 60,000 &295 & CL,GG  &1 \nl
6434 & 12/03/2001 & McD 2.7-m & 3700-10000 & 60,000 &450 & CL,GG  &1 \nl
114783 & 12/03/2001 & McD 2.7-m & 3700-10000 & 60,000 &225 & CL,GG  &2 \nl
202206 & 12/03/2001 & McD 2.7-m & 3700-10000 & 60,000 &320 & CL,GG  &16 \nl
141937 & 03/25/2002 & McD 2.7-m & 3700-10000 & 59,000 &355 & CL &16 \nl
136118 & 03/25/2002 & McD 2.7-m & 3700-10000 & 59,000 &385 & CL &17 \nl
168746 & 03/25/2002 & McD 2.7-m & 3700-10000 & 59,000 &230 & CL &9 \nl
\enddata
\tablenotetext{a}{The S/N ratio corresponds to the value near 6700 \AA.}
\tablenotetext{b}{The observer codes correspond to the following observers: 
CL, Chris Laws; GG, Guillermo Gonzalez; NS, Nicholas Suntzeff}
\tablenotetext{c}{The discoverer codes correspond to the following studies: 
1, Unpublished, Geneva Group Website; 2, Vogt et al. (2001);
3, Hatzes et al. (2000); 4, Sivan et al. (2000); 5, Fischer et al. (1999);
6, Marcy et al. (2000); 7, Vogt et al. (2000); 8, Santos et al. (2001b);
9, Pepe et al. (2002); 10, Jones et al. (2002); 11, Naef et al. (2001);
12, Udry et al. (2000); 13, Unpublished, Berkely Group Website; 14, 
Butler \& Marcy (1996); 15 Marcy \& Butler (1996); 16, Udry et al. 
(2002); 17, Fischer et al. (2002).}

\end{deluxetable}

\clearpage

\begin{deluxetable}{lcccccccc}
\tablecaption{Measured Equivalent Widths for McDonald 2.7m Data}
\scriptsize
\rotate
\tablewidth{0pt}
\tablehead{
\colhead{Species} & \colhead{$\lambda_{\rm o}$(\AA)} & 
\colhead{HD\,} & \colhead{HD\,} & \colhead{HD\,} & 
\colhead{HD\,} & \colhead{HD\,} & \colhead{HD\,} & 
\colhead{HD\,}
}
\startdata
These tables are available upon the request.
\enddata

\end{deluxetable}

\clearpage

\begin{deluxetable}{lcccccccc}
\tablecaption{Measured Equivalent Widths for CTIO 4m Data}
\scriptsize
\rotate
\tablewidth{0pt}
\tablehead{
\colhead{Species} & \colhead{$\lambda_{\rm o}$(\AA)} & 
\colhead{HD\,} & \colhead{HD\,} & \colhead{HD\,} & 
\colhead{HD\,} & \colhead{HD\,} & \colhead{HD\,} & 
\colhead{HD\,}
}
\startdata
These tables are available on request.
\enddata

\end{deluxetable}

\clearpage

\begin{deluxetable}{lccccc}
\tablecaption{Spectroscopically-Determined Physical Parameters of all Program Stars}
\scriptsize
\setlength{\tabcolsep}{0.06in}
\tablewidth{0pt}
\tablehead{
\colhead{Star} & \colhead{T$_{\rm eff}$} & \colhead{$\log g$} & 
\colhead{$\xi_{\rm t}$} & \colhead{[Fe/H]} & \colhead{N(Fe I,II)}\\
\colhead{(HD)} & \colhead{(K)} & \colhead{} & 
\colhead{(km~s$^{\rm -1}$)} & \colhead{} & \colhead{}
}
\startdata
4203 & $5587 \pm 54$ & $4.15 \pm 0.08$ & $1.05 \pm 0.06$ & $+0.40 \pm 0.04$ & 63, 9\nl
4208 & $5586 \pm 34$ & $4.39 \pm 0.04$ & $0.69 \pm 0.10$ & $-0.25 \pm 0.03$ & 56, 8\nl
6434 & $5705 \pm 84$ & $4.21 \pm 0.10$ & $0.67 \pm 0.13$ & $-0.55 \pm 0.07$ & 61, 8\nl
8574 & $6034 \pm 33$ & $4.19 \pm 0.11$ & $1.10 \pm 0.07$ & $+0.02 \pm 0.03$ & 66, 9\nl
16141\tablenotemark{a} & $5801 \pm 31$ & $4.24 \pm 0.04$ & $1.01 \pm 0.05$ & $+0.19 \pm 0.03$ & 60, 8\nl
19994 & $6164 \pm 47$ & $4.22 \pm 0.07$ & $1.82 \pm 0.11$ & $+0.14 \pm 0.04$ & 65, 9\nl
22049 & $5086 \pm 50$ & $4.41 \pm 0.11$ & $0.90 \pm 0.09$ & $-0.09 \pm 0.03$ & 61, 8\nl
27442 & $4797 \pm 101$ & $3.27 \pm 0.23$ & $1.20 \pm 0.13$ & $+0.41 \pm 0.05$ & 48, 8\nl
28185 & $5670 \pm 30$ & $4.54 \pm 0.05$ & $0.94 \pm 0.06$ & $+0.24 \pm 0.02$ & 65, 9\nl
33636 & $5930 \pm 33$ & $4.29 \pm 0.07$ & $1.01 \pm 0.10$ & $-0.11 \pm 0.03$ & 58, 8\nl
37124\tablenotemark{a} & $5551 \pm 34$ & $4.43 \pm 0.07$ & $0.60 \pm 0.18$ & $-0.37 \pm 0.03$ & 59, 8\nl
46375\tablenotemark{a} & $5241 \pm 44$ & $4.41 \pm 0.09$ & $0.69 \pm 0.11$ & $+0.30 \pm 0.03$ & 60, 8\nl
50554 & $5984 \pm 31$ & $4.37 \pm 0.05$ & $1.04 \pm 0.09$ & $+0.02 \pm 0.02$ & 64, 9\nl
68988 & $5922 \pm 57$ & $4.40 \pm 0.06$ & $1.08 \pm 0.07$ & $+0.34 \pm 0.04$ & 65, 9\nl
82943 & $6008 \pm 34$ & $4.43 \pm 0.06$ & $1.01 \pm 0.07$ & $+0.26 \pm 0.03$ & 56, 8\nl
83443 & $5389 \pm 66$ & $4.36 \pm 0.13$ & $0.81 \pm 0.12$ & $+0.36 \pm 0.04$ & 57, 8\nl
95128\tablenotemark{b} & $5861 \pm 30$ & $4.29 \pm 0.06$ & $1.01 \pm 0.07$ & $+0.05 \pm 0.02$ & 66, 9\nl
106252 & $5852 \pm 31$ & $4.39 \pm 0.05$ & $0.97 \pm 0.09$ & $-0.05 \pm 0.03$ & 64, 8\nl
108147 & $6316 \pm 91$ & $4.58 \pm 0.15$ & $0.99 \pm 0.18$ & $+0.23 \pm 0.06$ & 56, 8\nl
114783 & $5130 \pm 43$ & $4.50 \pm 0.09$ & $0.94 \pm 0.08$ & $+0.17 \pm 0.02$ & 66, 8\nl
117176\tablenotemark{b} & $5530 \pm 45$ & $3.99 \pm 0.05$ & $1.05 \pm 0.05$ & $-0.02 \pm 0.04$ & 66, 9\nl
121504 & $5941 \pm 74$ & $4.37 \pm 0.08$ & $1.00 \pm 0.18$ & $+0.12 \pm 0.05$ & 55, 8\nl
136118 & $6231 \pm 50$ & $4.29 \pm 0.08$ & $1.96 \pm 0.22$ & $-0.05 \pm 0.03$ & 62, 8\nl
141937 & $5856 \pm 49$ & $4.44 \pm 0.08$ & $0.96 \pm 0.06$ & $+0.14 \pm 0.04$ & 67, 9\nl
160691 & $5811 \pm 45$ & $4.42 \pm 0.06$ & $1.07 \pm 0.08$ & $+0.28 \pm 0.03$ & 48, 5\nl
168746 & $5577 \pm 44$ & $4.38 \pm 0.05$ & $0.93 \pm 0.06$ & $-0.06 \pm 0.03$ & 66, 8\nl
169830 & $6312 \pm 50$ & $4.15 \pm 0.06$ & $1.26 \pm 0.13$ & $+0.17 \pm 0.04$ & 48, 6\nl
190228 & $5276 \pm 36$ & $3.51 \pm 0.06$ & $0.99 \pm 0.05$ & $-0.24 \pm 0.03$ & 61, 8\nl
195019A & $5734 \pm 32$ & $4.09 \pm 0.10$ & $1.10 \pm 0.05$ & $+0.03 \pm 0.03$ & 60, 7\nl
202206 & $5716 \pm 38$ & $4.43 \pm 0.06$ & $0.98 \pm 0.06$ & $+0.33 \pm 0.03$ & 66, 9\nl
213240 & $6086 \pm 83$ & $4.51 \pm 0.16$ & $1.00 \pm 0.13$ & $+0.23 \pm 0.06$ & 51, 8\nl

\enddata

\tablenotetext{a}{These stars were previously analyzed using more limited Keck spectra in Paper VI.}
\tablenotetext{b}{These stars were previously analyzed using more limited McDonald 2.1m spectra in Paper II.}

\end{deluxetable}

\clearpage

\begin{deluxetable}{lccccccc}
\tablecaption{Derived Parameters of all Program Stars}
\scriptsize
\setlength{\tabcolsep}{0.04in}
\tablewidth{0pt}
\tablehead{
\colhead{Star} & \colhead{M$_{\rm V}$\tablenotemark{a}} & 
\colhead{Age\tablenotemark{b}} & \colhead{Mass\tablenotemark{b}} & 
\colhead{$\log g_{\rm evol}$\tablenotemark{b}} & 
\colhead{U, V, W\tablenotemark{c}} & 
\colhead{$\log R^{'}_{\rm HK}$\tablenotemark{d}} & 
\colhead{Age\tablenotemark{e}}\\
\colhead{(HD)} & \colhead{} & \colhead{(Gyr)} & \colhead{(M$_{\odot}$)} & 
\colhead{} & \colhead{(km~s$^{\rm -1}$)} & \colhead{} & \colhead{(Gyr)}
}
\startdata
4203 & $4.24 \pm 0.21$ & $7.6 \pm 0.7$ & $1.05 \pm 0.03$ & $4.15 \pm 0.10$ & $-17.9$,$-36.3$,$+2.7$ & $-5.13$ &8.7\nl
4208 & $5.22 \pm 0.08$ & $11.7 \pm 0.8$ & $0.84 \pm 0.02$ & $4.41 \pm 0.03$ & $-42.5$,$+1.1$,$-51.8$ & $-4.93$ &4.3 \nl
6434 & $4.69 \pm 0.08$ & $> 17 $ & $0.80 \pm 0.04$ & $4.25 \pm 0.05$ & $+95.2$,$-6.1$,$+3.4$ & $-4.89$ &3.7 \nl
8574 & $3.90 \pm 0.16$ & see text & see text & see text & $-34.1$,$-30.7$,$-24.9$ & \nodata & \nodata \nl
16141 & $4.05 \pm 0.11$ & $3.2 \pm 0.2$ & $1.18 \pm 0.02$ & $4.20 \pm 0.03$ & $+94.5$,$-35.2$,$+7.5$ & $-5.05$ &6.7 \nl
19994 & $3.32 \pm 0.04$ & $2.9 \pm 0.1$ & $1.34 \pm 0.02$ & $4.09 \pm 0.03$ & $-10.5$,$-13.5$,$-0.9$ & $-4.84$ &3.1 \nl
22049 & $6.19 \pm 0.01$ & $2.0^{+3.0}_{-2.0}$ & $0.81 \pm 0.03$ & $4.58 \pm 0.04$ & $+6.4$,$+13.1$,$-14.6$ & $-4.47$ &0.7 \nl
27442 & $3.14 \pm 0.02$ & $3.2 \pm 0.2$ & $1.40 \pm 0.07$ & $3.42 \pm 0.09$ & $+25.1$,$-16.1$,$-13.2$ & \nodata &\nodata \nl
28185 & $4.81 \pm 0.09$ & $4.0 \pm 1.0$ & $1.03 \pm 0.02$ & $4.20 \pm 0.02$ & $-23.6$,$-28.8$,$-17.4$ & $-4.82$ &2.9 \nl
33636 & $4.71 \pm 0.08$ & $2.5 \pm 1.5$ & $1.03 \pm 0.02$ & $4.42 \pm 0.03$ & $+9.5$,$-24.2$,$+14.8$ & $-4.81$ &2.8 \nl
37124 & $5.07 \pm 0.08$ & $>18$ & $0.77 \pm 0.02$ & $4.33 \pm 0.15$ & $+42.7$,$-40.1$,$-37.3$ & $-4.90$ &3.9 \nl
46375 & $5.29 \pm 0.08$ & $16.5 \pm 3.5$ & $0.87 \pm 0.02$ & $4.34 \pm 0.10$ & $+20.3$,$-13.4$,$+14.9$ & $-4.94$ &4.5 \nl
50554 & $4.38 \pm 0.07$ & $3.9 \pm 0.7$ & $1.08 \pm 0.02$ & $4.36 \pm 0.02$ & $+13.7$,$-4.0$,$-5.5$ & $-4.94$ &4.5 \nl
68988 & $4.35 \pm 0.12$ & $1.6 \pm 1.5 $ & $1.18 \pm 0.04$ & $4.37 \pm 0.05$ & $+85.0$,$-15.8$,$-3.8$ & $-5.07$ &7.1 \nl
82943 & $4.35 \pm 0.05$ & $0.6 \pm 0.5$ & $1.18^{+0.05}_{-0.01}$ & $4.39 \pm 0.02$ & $+20.2$,$-13.9$,$-2.9$ & $-4.95$ &4.7 \nl
83443 & $5.04 \pm 0.08$ & $10.7 \pm 1.3$ & $0.93 \pm 0.01$ & $4.31 \pm 0.10$ & $+29.9$,$-24.7$,$-6.0$ & $-4.85$ &3.2 \nl
95128 & $4.29 \pm 0.03$ & $6.8 \pm 0.3$ & $1.05 \pm 0.01$ & $4.25 \pm 0.02$ & $-14.0$,$+3.6$,$+6.6$ & $-5.26$ &12.7 \nl
106252 & $4.54 \pm 0.07$ & $6.8 \pm 0.5$ & $1.00 \pm 0.01$ & $4.34 \pm 0.03$ & $+38.5$,$-37.6$,$+6.5$ & $-4.97$ &5.0 \nl
108147 & $4.06 \pm 0.06$ & $<1.0$ & $1.23 \pm 0.03$ & $4.36 \pm 0.04$ & $-20.3$,$-5.4$,$-8.3$ & $-4.72$ &2.0 \nl
114783 & $6.02 \pm 0.05$ & $<4.5$ & $0.88 \pm 0.03$ & $4.54 \pm 0.03$ & $-5.4$,$+3.3$,$-2.9$ & $-4.96$ &4.8 \nl
117176 & $3.68 \pm 0.03$ & $7.9 \pm 0.4$ & $1.05 \pm 0.02$ & $3.92 \pm 0.05$ & $+23.2$,$-45.9$,$+2.2$ & $-4.74$ & see text \nl
121504 & $4.30 \pm 0.09$ & $4.0^{+1.0}_{-1.5}$ & $1.11 \pm 0.02$ & $4.30 \pm 0.08$ & $-17.8$,$-45.9$,$+4.0$ & $-4.81$ &2.8 \nl
136118 & $3.34 \pm 0.09$ & $3.2 \pm 0.2$ & $1.28 \pm 0.01$ & $4.08 \pm 0.04$ & $-10.9$,$-10.2$,$+23.0$ & $-4.88$ &3.6 \nl
141937 & $4.63 \pm 0.08$ & $2.5^{+1.0}_{-1.5}$ & $1.08 \pm 0.02$ & $4.41 \pm 0.02$ & $+13.0$,$+19.3$,$-2.6$ & $-4.65$ &1.6 \nl
160691 & $4.20 \pm 0.03$ & $4.5 \pm 0.4$ & $1.14 \pm 0.01$ & $4.25 \pm 0.04$ & $-3.7$,$-2.4$,$+1.9$ & $-5.02$ &6.0 \nl
168746 & $4.78 \pm 0.09$ & $14.8 \pm 1.0$ & $0.89 \pm 0.02$ & $4.21 \pm 0.10$ & $-9.4$,$-16.1$,$+2.8$ &\nodata &\nodata \nl
169830 & $3.10 \pm 0.07$ & $2.3 \pm 0.2$ & $1.41 \pm 0.02$ & $4.07 \pm 0.02$ & $-7.1$,$+2.5$,$+7.5$ & $-4.93$ &4.3 \nl
190228 & $3.33 \pm 0.11$ & $4.5 \pm 0.2$ & $1.23 \pm 0.03$ & $3.68 \pm 0.10$ & $-10.0$,$-41.0$,$-29.9$ &\nodata &\nodata \nl
195019A & $4.01 \pm 0.07$ & $8.7 \pm 0.4$ & $1.04 \pm 0.01$ & $4.09 \pm 0.05$ & $-62.0$,$-69.7$,$-31.4$ & $-4.85$ &3.2 \nl
202206 & $4.75 \pm 0.11$ & $1.7 \pm 1.5$ & $1.10 \pm 0.05$ & $4.41 \pm 0.02$ & $+32.5$,$-13.3$,$-3.9$ &\nodata &\nodata \nl
213240 & $3.76 \pm 0.07$ & $2.9 \pm 0.4$ & $1.25 \pm 0.01$ & $4.21 \pm 0.02$ & $+35.1$,$-24.2$,$+29.3$ & $-4.80$ &2.7 \nl
\nl
1237 & $5.36 \pm 0.02$ & $0.7 \pm 0.7$ & $0.99 \pm 0.02$ & $4.51 \pm 0.01$ & $-23.1$,$-10.4$,$+8.4$ & $-4.27$ & $ <0.1 $ \nl
9826 & $3.45 \pm 0.02$ & $3.1 \pm 0.3$ & $1.30 \pm 0.01$ & $4.11 \pm 0.04$ & $+38.8$,$-16.5$,$-8.2$ & $-4.97$ &5.0 \nl
10697 & $3.71 \pm 0.06$ & $6.9 \pm 0.4$ & $1.12 \pm 0.02$ & $3.98 \pm 0.05$ & $+46.7$,$-22.2$,$+23.0$ & $-5.02$ & see text \nl
12661 & $4.58 \pm 0.07$ & $3.6 \pm 0.8$ & $1.08 \pm 0.01$ & $4.34 \pm 0.03$ & $+61.7$,$-23.5$,$+3.7$ & $-5.12$ &8.4 \nl
17051 & $4.22 \pm 0.02$ & $0.5 \pm 0.5$ & $1.20 \pm 0.02$ & $4.38 \pm 0.02$ & $-21.2$,$-11.4$,$-2.8$ & $-4.65$ &1.6 \nl
38529 & $2.81 \pm 0.08$ & $3.0 \pm 0.2$ & $1.42 \pm 0.05$ & $3.72 \pm 0.03$ & $-3.5$,$-19.1$,$-28.2$ & $-4.89$ & see text \nl
52265 & $4.05 \pm 0.05$ & $0.5^{+0.5}_{-0.3}$ & $1.25 \pm 0.01$ & $4.35 \pm 0.02$ & $-42.3$,$-14.7$,$-3.2$ & $-4.91$ &4.0 \nl
75289 & $4.04 \pm 0.03$ & $0.8^{+0.5}_{-0.8}$ & $1.24^{+0.03}_{-0.01}$ & $4.33 \pm 0.02$ & $+30.9$,$-6.6$,$-15.7$ & $-4.96$ &4.8 \nl
75732 & $5.47 \pm 0.02$ & $11.0 \pm 5.0$ & $0.89 \pm 0.06$ & $4.40 \pm 0.03$ & $-27.2$,$-12.0$,$-2.1$ & $-4.97$ &5.0 \nl
89744 & $2.78 \pm 0.06$ & $1.9 \pm 0.1$ & $1.53 \pm 0.02$ & $3.99 \pm 0.02$ & $-1.2$,$-23.6$,$-7.4$ & $-5.12$ & see text \nl
92788 & $4.76 \pm 0.07$ & $0.4^{+0.4}_{-0.2}$ & $1.12 \pm 0.03$ & $4.43 \pm 0.02$ & $+26.1$,$-16.3$,$-14.8$ & $-5.04$ &6.4 \nl
114762 & $4.26 \pm 0.12$ & $14.0 \pm 2.0$ & $0.91 \pm 0.03$ & $4.19 \pm 0.07$ & $-73.2$,$-62.9$,$+63.8$ &\nodata &\nodata \nl
120136 & $3.53 \pm 0.02$ & $0.7 \pm 0.6$ & $1.37 \pm 0.02$ & $4.26 \pm 0.03$ & $-23.7$,$-12.9$,$-1.6$ & $-4.73$ &2.1 \nl
130322 & $5.67 \pm 0.10$ & $<1.0$ & $0.91 \pm 0.03$ & $4.57 \pm 0.03$ & $+0.6$,$-20.0$,$-4.9$ & $-4.39$ &0.4 \nl
134987 & $4.42 \pm 0.05$ & $5.1^{+1.4}_{-0.5}$ & $1.10 \pm 0.02$ & $4.29 \pm 0.03$ & $-10.6$,$-34.1$,$+26.7$ & $-5.01$ &5.8 \nl
143761 & $4.18 \pm 0.03$ & $12.6 \pm 0.2$ & $0.90 \pm 0.01$ & $4.12 \pm 0.05$ & $+64.3$,$-29.8$,$+26.8$ & \nodata & \nodata \nl
145675 & $5.32 \pm 0.02$ & $9.8^{+7.1}_{-4.8}$ & $0.87 \pm 0.05$ & $4.36 \pm 0.06$ & $+33.9$,$-6.1$,$-10.1$ & $-5.07$ &7.1 \nl
168443 & $4.03 \pm 0.07$ & $9.3 \pm 0.6$ & $1.03 \pm 0.02$ & $4.02 \pm 0.04$ & $-19.8$,$-51.7$,$-0.7$ & $-5.08$ &7.4 \nl
177830 & $3.32 \pm 0.10$ & $3.5 \pm 0.4$ & $1.38 \pm 0.05$ & $3.55 \pm 0.06$ & $-13.2$,$-64.4$,$-1.2$ & $-5.28$ & see text \nl
186427 & $4.60 \pm 0.02$ & $9.1 \pm 0.8$ & $0.98 \pm 0.01$ & $4.30 \pm 0.04$ & $+27.6$,$-24.4$,$+4.1$ &\nodata &\nodata \nl
187123 & $4.43 \pm 0.07$ & $5.0 \pm 0.6$ & $1.07 \pm 0.01$ & $4.32 \pm 0.03$ & $+12.6$,$-9.3$,$-37.4$ &\nodata &\nodata \nl
192263\tablenotemark{g} & $6.30 \pm 0.05$ & see text & $\sim 0.76$ & $\sim 4.54$ & $-6.1$,$+16.9$,$+25.7$ & $-4.37$ &0.3 \nl
209458 & $4.29 \pm 0.10$ & $2.4^{+1.1}_{-1.9}$ & $1.12 \pm 0.02$ & $4.36 \pm 0.04$ & $+4.3$,$-9.6$,$+6.6$ & $-4.93$ &4.3 \nl
210277 & $4.90 \pm 0.04$ & $8.1^{+1.9}_{-2.9}$ & $0.98 \pm 0.03$ & $4.35 \pm 0.02$ & $+14.0$,$-44.2$,$-0.3$ & $-5.06$ &6.9 \nl
217014 & $4.52 \pm 0.02$ & $4.3 \pm 1.0$ & $1.07 \pm 0.01$ & $4.35 \pm 0.02$ & $-5.2$,$-23.6$,$+21.6$ & $-5.07$ &7.1 \nl
217107 & $4.70 \pm 0.03$ & $6.3 \pm 1.3$ & $1.02 \pm 0.02$ & $4.31 \pm 0.02$ & $+8.5$,$-2.5$+16.4,$$ & $-5.00$ &5.6 \nl
222582 & $4.57 \pm 0.10$ & $8.7 \pm 0.8$ & $0.98 \pm 0.02$ & $4.32 \pm 0.03$ & $+46.7$,$+5.4$,$-5.1$ & $-5.00$ &5.6 \nl

\enddata
\tablenotetext{a}{Calculated from the {\it Hipparcos} parallaxes.}
\tablenotetext{b}{Derived from Padova Stellar Isochrones (Salasnich et al. 2000).}
\tablenotetext{c}{Space velocities are relative to the Local 
Standard of Rest (LSR).  The assumed solar motion with respect to the LSR 
is (U, V, W) $= (10,6,6)$ km~s$^{\rm -1}$.  Positive U directed toward the 
Galactic center.}
\tablenotetext{d}{R$^{'}_{\rm HK}$ values are taken from the corresponding 
discovery papers listed in the Introduction, except HD75332 (Saar \& Brandenburg, 1999),
HD83443 (Butler et al., 2002), and HD22049 (Henry et al., 1996).}
\tablenotetext{e}{Calculated from the R$^{'}_{\rm HK}$ values and Henry 
et al.'s (1996) equation, which is taken originally from Donahue (1993).}
\tablenotetext{f}{HD75332 is not known to harbor a planet, but is included in this
table for a complete comparison with Table 14 of Paper VI.}
\tablenotetext{g}{The existence of the planet orbiting HD192263 has
recently been questioned by Henry et al. (2002).}
\end{deluxetable}


\clearpage

\begin{thebibliography}{}
\parsep 0pt
\itemsep -3pt

\bibitem[Prieto \& Lambert 1999]{APL:99}
Allende Prieto, C. \& Lambert, D. 1999, \aap, 352, 555

\bibitem[Barbieri \& Gratton 2002]{BarG:02}
Barbieri, M. \& Gratton, R. G. 2002, \aap, 384, 879

\bibitem[Barnes 2001]{Bar:01}
Barnes, S. 2001, \aj, 561, 1095

\bibitem[Bertelli {et~al.} 1994]{Chen:94}
Bertelli, G., Bressan, A., Chiosi, C., Fagotto, F. \& Nasi E. 1994, \aap, 106S, 275

\bibitem[Bouchy {et~al.} 2001]{Bou:01}
Bouchy, F., Pepe, F. \& Queloz, D. 1999, \aap, 374, 733

\bibitem[Butler \& Marcy 1996]{Butl:96}
Butler, R. P. \& Marcy, G. W. 1996, \apj, 464, L153

\bibitem[Butler 2000]{Butl:00}
Butler, R. P., Vogt, S. S., Marcy, G. W., Fischer, D. A., Henry, G. W., \& 
Apps, K. 2000, \apj, in press

\bibitem[Cochran {et~al.} 2002]{Coch:02}
Cochran, W. D., Hatzes, A. P. \& Paulson, D. B. 2002, \aj, 124, 565

\bibitem[Cumming {et~al.} 1999]{Cumm:99}
Cumming, A., Marcy, G. W. \& Butler, R. P. 1999, \apj, 526, 890

\bibitem[Donahue 1993]{Dona:93}
Donahue, R. A. 1993, Ph.D. thesis, New Mexico State Univ.

\bibitem[Donahue 1998]{Dona:98}
Donahue, R. A. 1998, in ASP Conf. Ser. 154, 10th Cambridge Workshop
on Cool Stars, Stellar Systems, and the Sun, ed. R. A. Donahue \&
J. A. Bookbinder (San Francisxo:ASP), 1235

\bibitem[Edvardsson {et~al.} 1993]{Edva:93}
Edvardsson, B., Andersen, J., Gustafsson, B., Lambert, D. L., \& Nissen, 
P. E. et al. 1993, \aap, 275, 101

\bibitem[ESA 1997]{ESA:97}
ESA 1997, The {\it Hipparcos} and {\it Tycho} Catalogue, ESA SP-1200

\bibitem[Fischer {et~al.} 1999]{Fisc:99}
Fischer, D. A., Marcy, G. W., Butler, R. P., Vogt, S. S., \& Apps, K. 1999, 
\pasp, 111, 50

\bibitem[Fischer {et~al.} 2002]{Fisc:02}
Fischer, D. A., Marcy, G. W., Butler, R. P., Vogt, S. S., Walp, B., 
\& Apps, K. 2002, \pasp, 114, 529

\bibitem[Ford {et~al.} 1999]{Ford:99}
Ford, E. B., Rasio, F. A., \& Sills, A. 1999, \apj, 514, 411

\bibitem[Fuhrmann {et~al.} 1998]{Fuhr:98}
Fuhrmann, K., Pfeiffer, M. J. \& Bernkopf, J. 1998, \aap, 336, 942

\bibitem[Girardi {et~al.} 2000]{Gir:00}
Girardi, L., Bressan, A., Bertelli, G., \& Carraro, G. 2000, \aa, 141S, 371

\bibitem[Gonzalez 1997]{Gonz:97}
Gonzalez, G. 1997, \mnras, 285, 403 (Paper I)

\bibitem[Gonzalez 1998]{Gonz:98}
Gonzalez, G. 1998, \aap, 334, 221 (Paper II)

\bibitem[Gonzalez 2003]{Gonz:03}
Gonzalez, G. 2003, RMP, 75, 101

\bibitem[Gonzalez \& Vanture 1998]{GVan:98}
Gonzalez, G. \& Vanture, A. D. 1998, \aap, 339, L29 (Paper III)

\bibitem[Gonzalez \& Laws 2000]{Gonz:00}
Gonzalez, G. \& Laws, C. 2000, \aj, 119, 390 (Paper V)

\bibitem[Gonzalez {et~al.} 1999]{GWS:99}
Gonzalez, G., Wallerstein, G., \& Saar, S. H. 1999, \apj, 511, L111 (Paper IV)

\bibitem[Gonzalez {et~al.} 2001]{G2plus:01}
Gonzalez, G., Laws, C., Tyagi, S., \& Reddy, B. E. 2001, \aj, 121, 432 (Paper VI)

\bibitem[Gratton {et~al.} 2001]{Grat:01}
Gratton, R. G., Bonnano, G., Claudi, R. U., Cosentino, R., 
Desidera, S., Lucatello, S., \& Scuderi, S. 2001, \aap, 377, 123

\bibitem[Hatzes {et~al.} 2000]{Hatz:00}
Hatzes, A., Cochran, W., McArthur, B., Baliunas, S. L., 
Walker, G., et al. 2000,  \apjl, 544, L145

\bibitem[Henry {et~al.} 1996]{Henr:96}
Henry, T. J., Soderblom, D. R., Donahue, R. A., \& Baliunas, S. L. 1996, 
\aj, 111, 439

\bibitem[Henry {et~al.} 2000]{Henr:00}
Henry, G. W., Butler, R. P., \& Vogt, S. S. 2000, \apj, 529, L41

\bibitem[Henry {et~al.} 2002]{Henr:02}
Henry, G. W., Donahue, R. A., \& Baliunas, S. L. 2002, \apj, 577, L111

\bibitem[Jones {et~al.} 2002]{Jone:02}
Jones, H. R. A., Butler, R. P., Tinney, C. G., Marcy, G. W., 
Penny, A. J. 2002, \mnras, in press

\bibitem[Kotoneva {et~al.} 2002]{Koto:02}
Kotoneva, E., Flynn, C., Jimenez, R. 2002, \mnras, 335, 1147

\bibitem[Laughlin 2000]{Laug:00}
Laughlin, G. 2000, \apj, 545, 1064L

\bibitem[Laws \& Gonzalez 2001]{LawG:01}
Laws, C. \& Gonzalez, G. 2001, \apj, 553, 405

\bibitem[Luhman \& Jayawardhana 2002]{Luh:02}
Luhman, K. L. \& Jayawardhana, Ray 2002, \apj, 566, 1132L

\bibitem[Maeder 1976]{Maed:76}
Maeder, A. 1976, \aap, 47, 389

\bibitem[Marcy \& Butler 1996]{Marc:96}
Marcy, G. \& Butler, R. P. 1996, \apj, 464, L147

\bibitem[Marcy {et~al.} 2000]{Marc:00}
Marcy, G. W., Butler, R. P., \& Vogt, S. S. 2000, \apj, 536, L43

\bibitem[Martell \& Laughlin 2002]{MarL:02}
Martell, S. \& Laughlin, G. 2002, \apj, 577L, 45

\bibitem[Murray \& Chaboyer 2002]{MurChab:02}
Murray, N. \& Chaboyer, B. 2002, \apj, 566, 442

\bibitem[Naef {et~al.} 2000a]{Naea:00}
Naef, D., Mayor, M., Pepe, F., Queloz, D., Udry, S., \& Burnet, M. 2000a, 
Disks, Planetesimals, and Planets, ed. F. Garzon, C. Eiroo, D. de Winter, 
\& T. J. Mahoney, ASP Conf. Ser (San Francisco: ASP), in press

\bibitem[Naef {et~al.} 2001]{Naef:01}
Naef, D., Mayor, M., Pepe, F., Queloz, D., Santos, N. et al. 2001, 
\aap, 375, 205

\bibitem[Nidever {et~al.} 2002]{Nide:02}
Nidever, D. L., Marcy, G. W., Butler, R. P., Fischer, D. A., 
Vogt, S. S. 2002, \apjs, 141, 503

\bibitem[Noyes {et~al.} 1984]{Noy:84}
Noyes, R. W., Hartmann, I., Baliunas, S. L., Duncan, D. K. \&
Vaughan, A. H. 1984, \apj, 279, 763

\bibitem[Pepe {et~al.} 2002]{Pepe:02}
Pepe, F., Mayor, M., Galland, F., Naef, D., Queloz, D., et al. 2002, 
\aap, 388, 632

\bibitem[Perrier {et~al.} 2002]{Perr:02}
Perrier, C. et al. 2002, \aap, in press.

\bibitem[Pinsonneault {et~al.} 2001]{Pin:01}
Pinsonneault, M. H., DePoy, D. L., \& Coffee, M. 2001, \apj, 556, L59

\bibitem[Reid 2002]{Reid:02}
Reid, I. N. 2002, \pasp, 114, 306

\bibitem[Saar \& Brandenburg 1999]{Saar:99}
Saar, S. H., \& Brandenburg, A. 1999, \apj, 524, 295

\bibitem[Salasnich {et~al.} 2000]{Sala:00}
Salasnich, B., Girardi, L., Weiss, A., \& Chiosi, C. 2000, \aap, 361, 1023S

\bibitem[Santos {et~al.} 2000]{San:00}
Santos, N. C., Israelian, G., \& Mayor, M. 2000, \aap, 363, 228

\bibitem[Santos {et~al.} 2001a]{Sana:01}
Santos, N. C., Israelian, G., \& Mayor, M. 2001, \aap, 373, 1019

\bibitem[Santos {et~al.} 2001b]{Sanb:01}
Santos, N. C., Mayor, M., Naef, D., Pepe, F., Queloz, D., Udry, S., \&
Burnet, M. 2001, \aap, 379, 999

\bibitem[Santos {et~al.} 2002]{Sanb:02}
Santos, N. C., Israelian, G., Mayor, M., Rebolo., R., \& Udry, S. 2002,
\aap, 398, 363

\bibitem[Schaerer {et~al.} 1993]{Scha:93}
Schaerer, D., Charbonnel, C., Meynet, G., Maeder, A., \& Schaller, G. 
1993, \aaps, 102, 339

\bibitem[Schaller {et~al.} 1992]{Schl:92}
Schaller, G., Schaerer, D., Meynet, G., \& Maeder, A. 1992, \aaps, 96, 269

\bibitem[Sivan {et~al.} 2000]{Siva:00}
Sivan, J. P. et al. 2000, IAU Symposium 202, in press

\bibitem[Smith {et~al.} 2001]{Sm:01}
Smith, V., Cunha, K., \& Lazzaro, D. 2001, \aj, 121, 320

\bibitem[Soderblom {et~al.} 1991]{Sod:91}
Soderblom, D., Duncan, D., Johnson, D. 1991, \apj, 375, 722

\bibitem[Takeda {et~al.} 2001]{Tak:01}
Takeda, Y., Sato, B., Kambe, E., Aoki, W., et al. 2001, \pasj, 53, 1211

\bibitem[Udry {et~al.} 2000]{Udry:00}
Udry, S., Mayor, M., \& Queloz, D. 2000b, IAU Symposium 202

\bibitem[Udry {et~al.} 2002]{Udry:02}
Udry, S., Mayor, M., Naef, D., Pepe, F., Queloz, D., Santos, N. C. et al. 
2002, \aap, 390, 26

\bibitem[Vogt {et~al.} 2000]{Vogta:00}
Vogt, S. S., Marcy, G. W., Butler, R. P., \& Apps, K. 2000, \apj, 536, 902

\bibitem[Vogt {et~al.} 2001]{Vogtb:01}
Vogt, S. S., Butler, R. P., Marcy, G. W., Fischer, D., Pourbaix, D. et 
al. 2001, \apj, 568, 362

\bibitem[Wetherill 1994]{Wet:94}
Wetherill, G. W. 1994, \apss, 212, 23

\bibitem[Zhao {et~al.} 2002]{Zhao:02}
Zhao, G., Chen, Y. Q., Qiu, H. M., \& Li, Z. W. 2002, \aj, 124, 2224

\bibitem[Zucker 2000]{Zuck:00}
Zucker, S. \& Mazeh, T. 2000, \apj, 531, L67

\end{thebibliography}
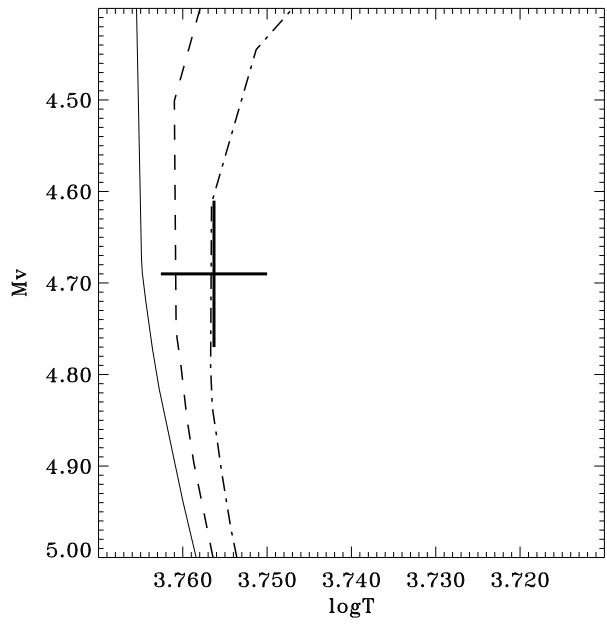
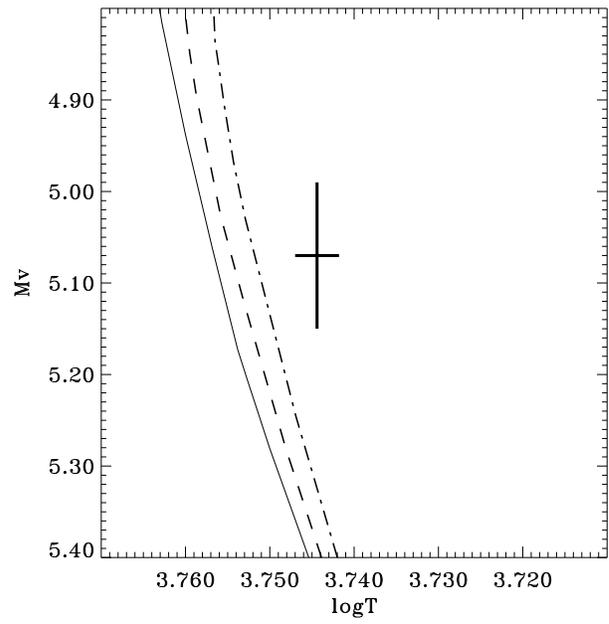
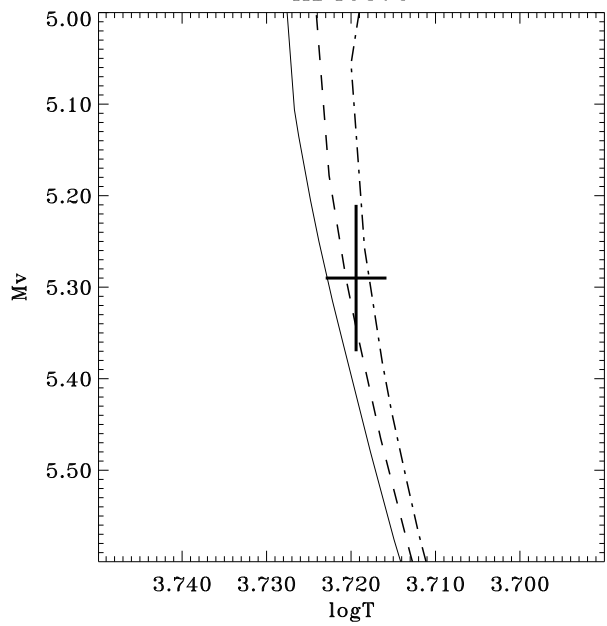
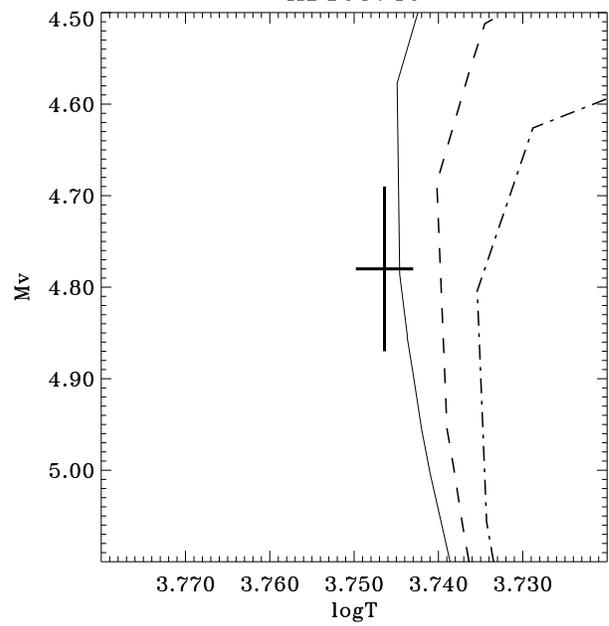
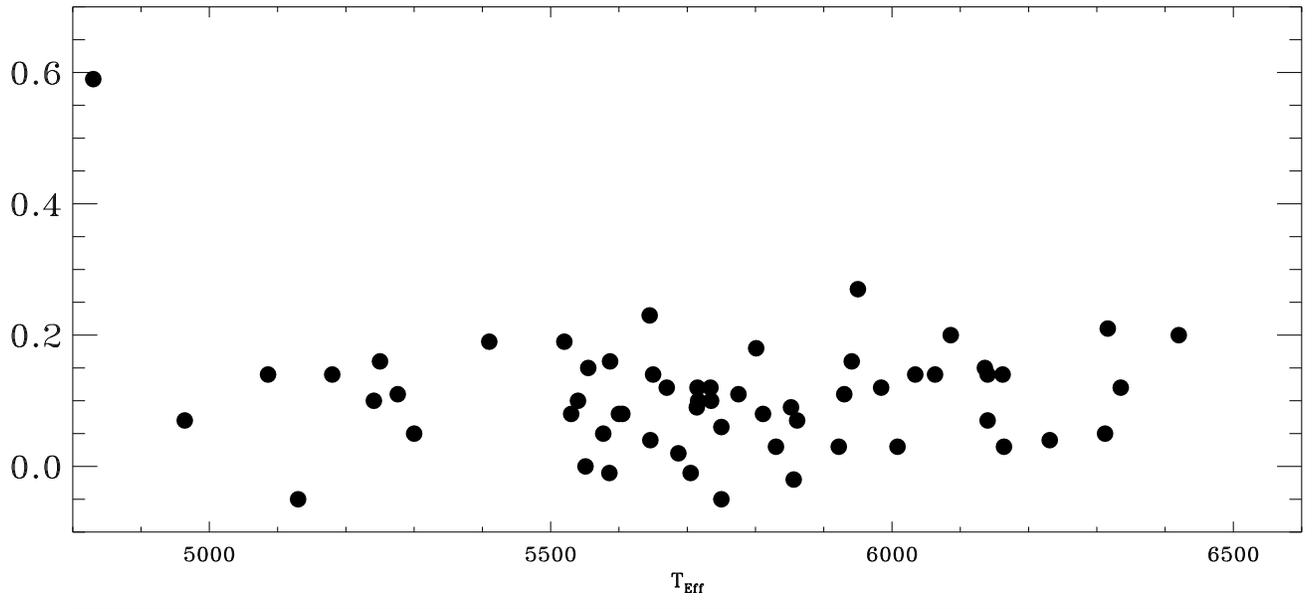
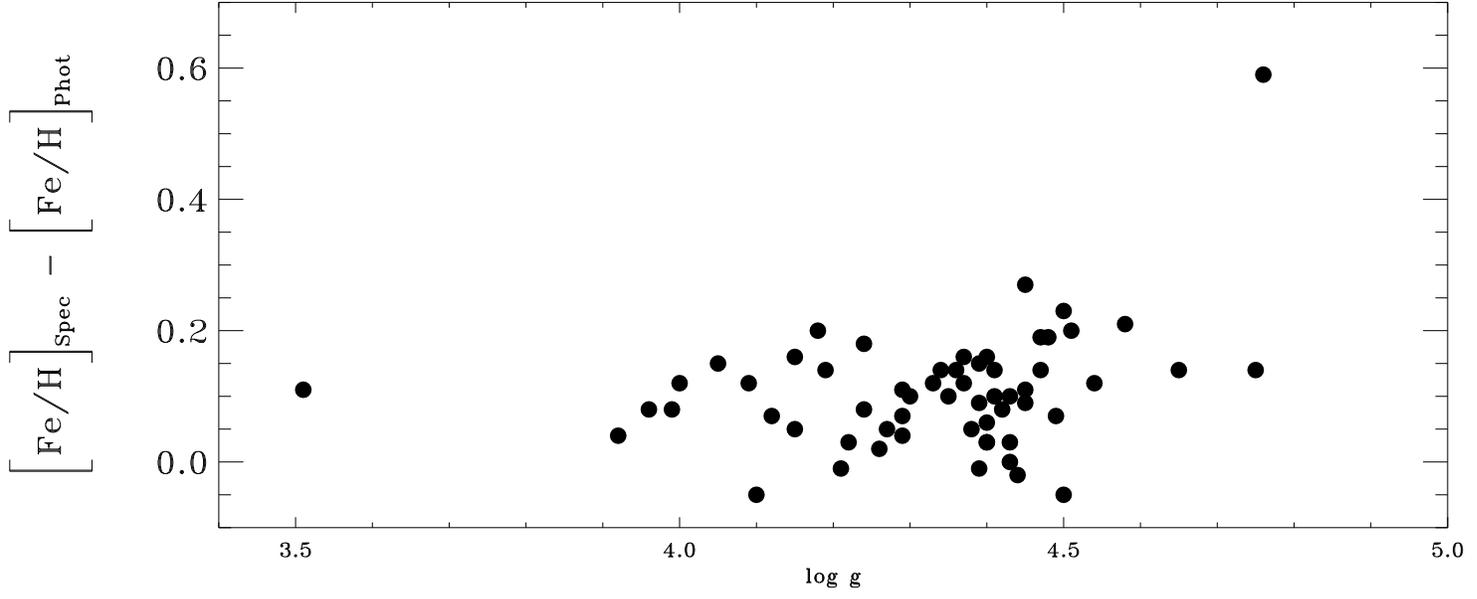
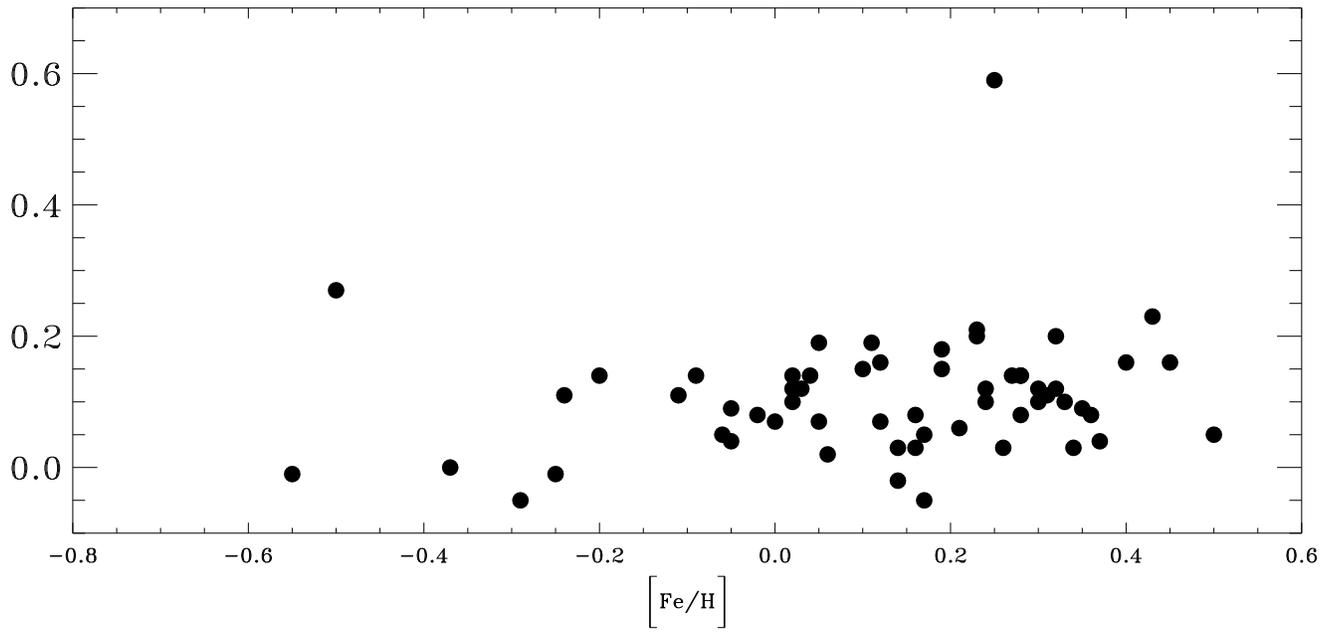
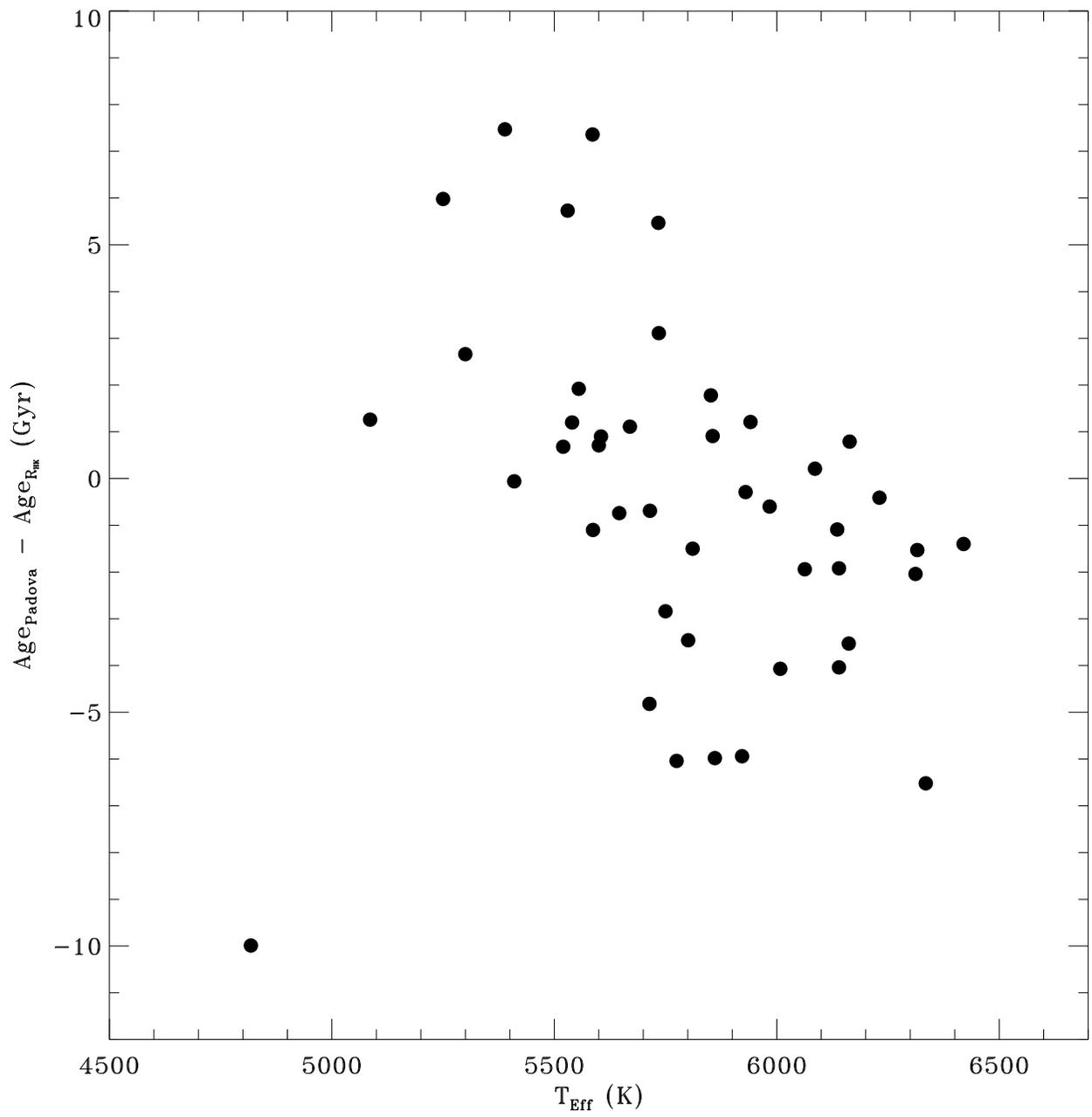
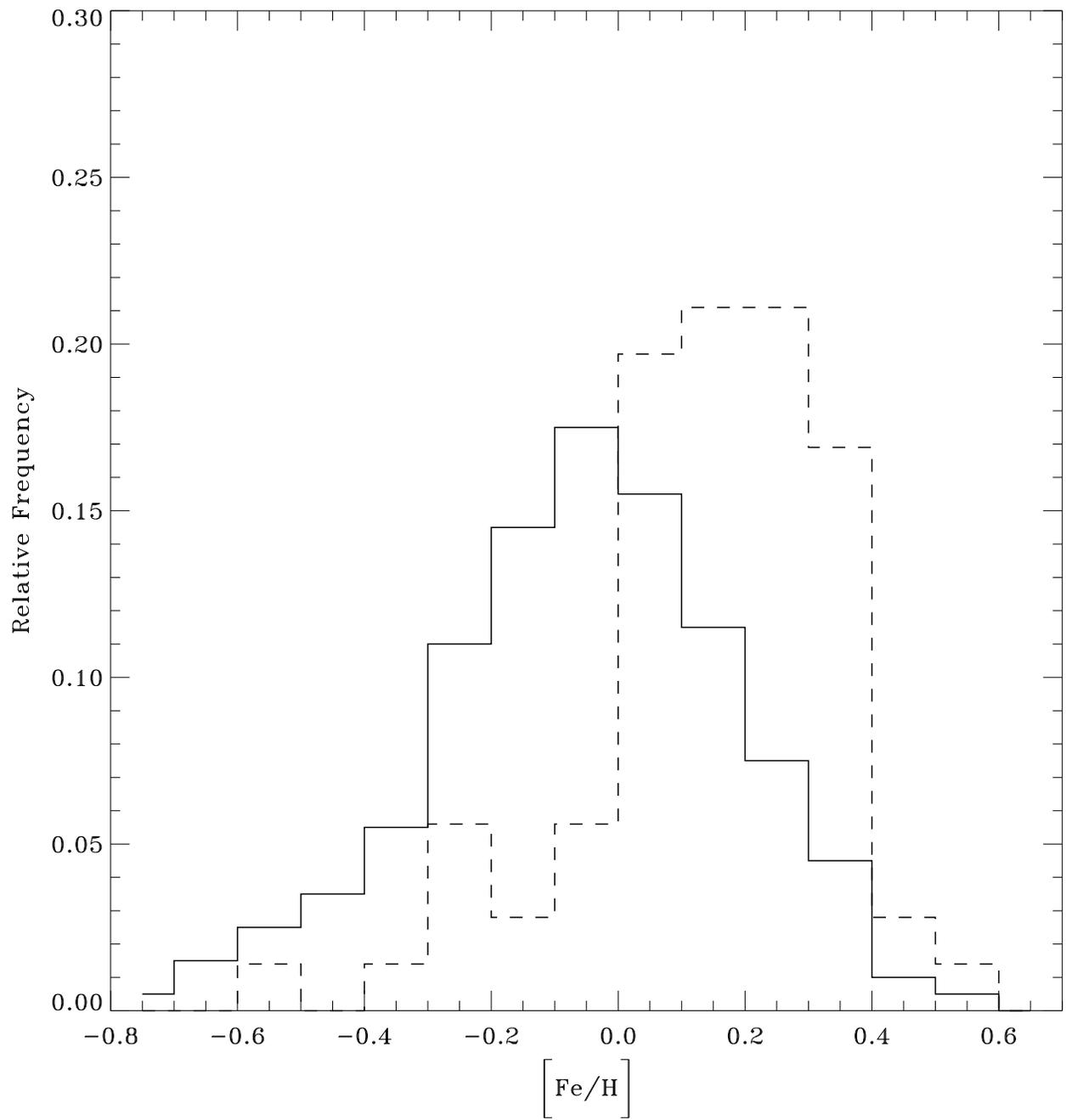
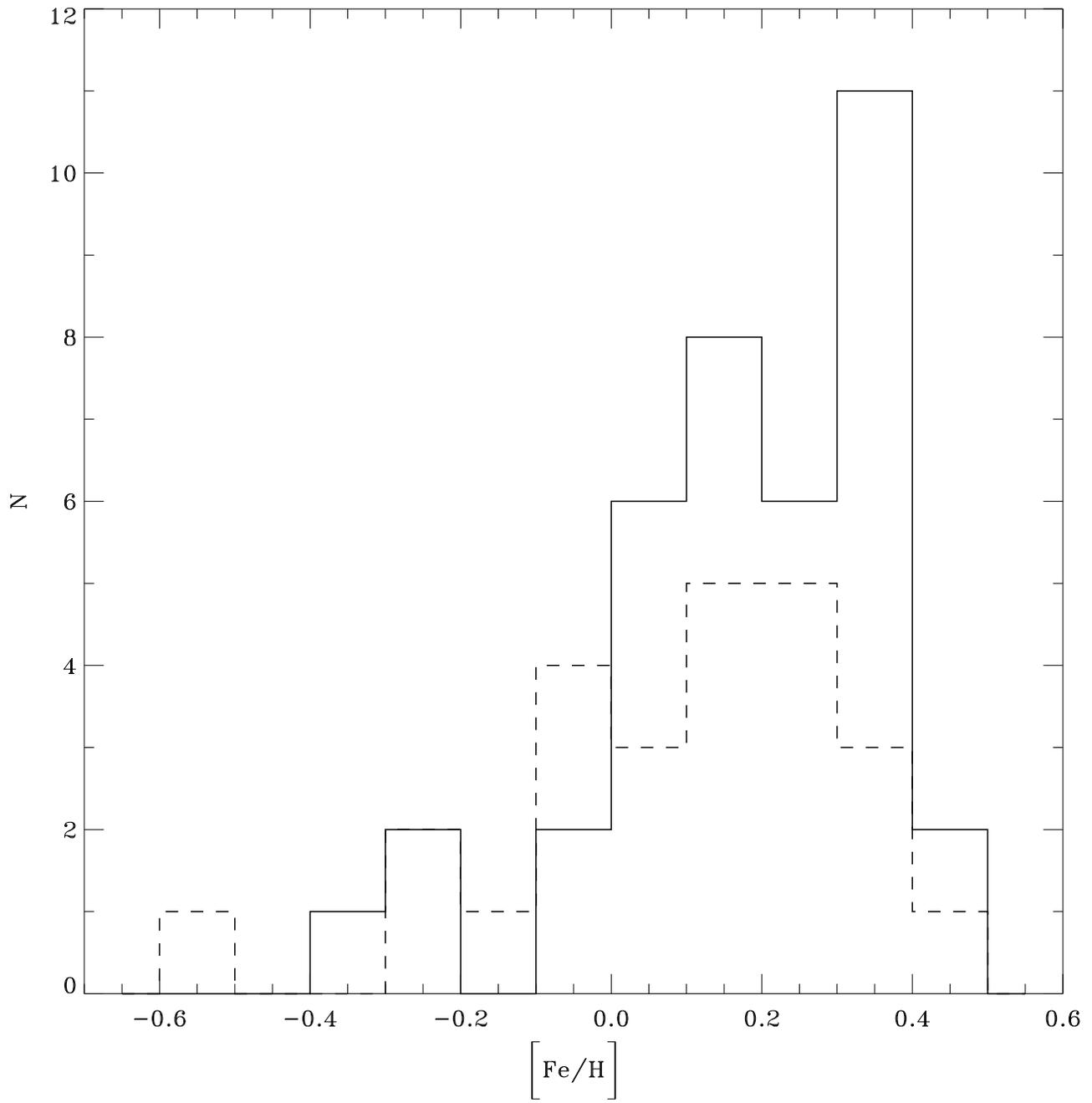
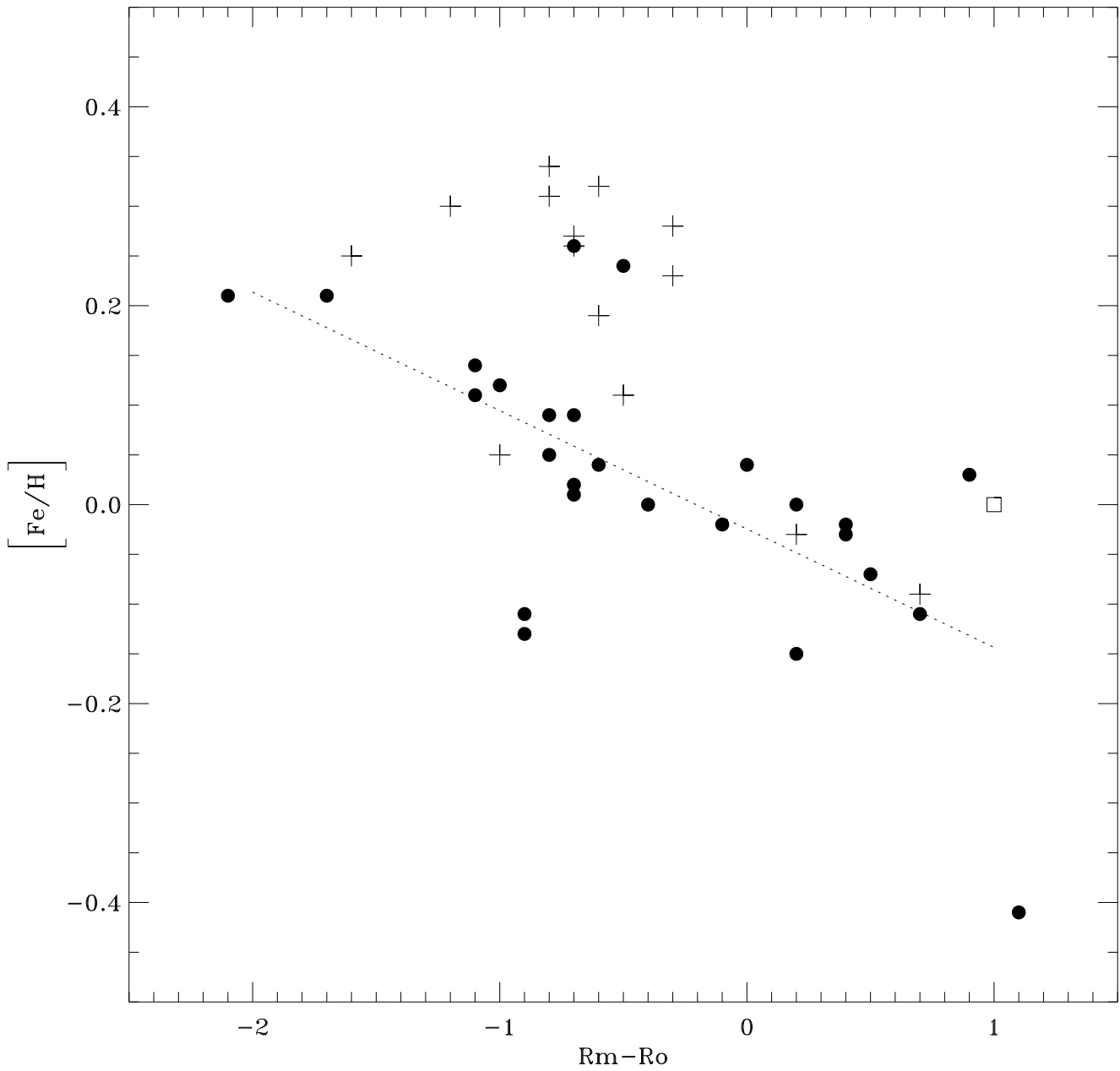
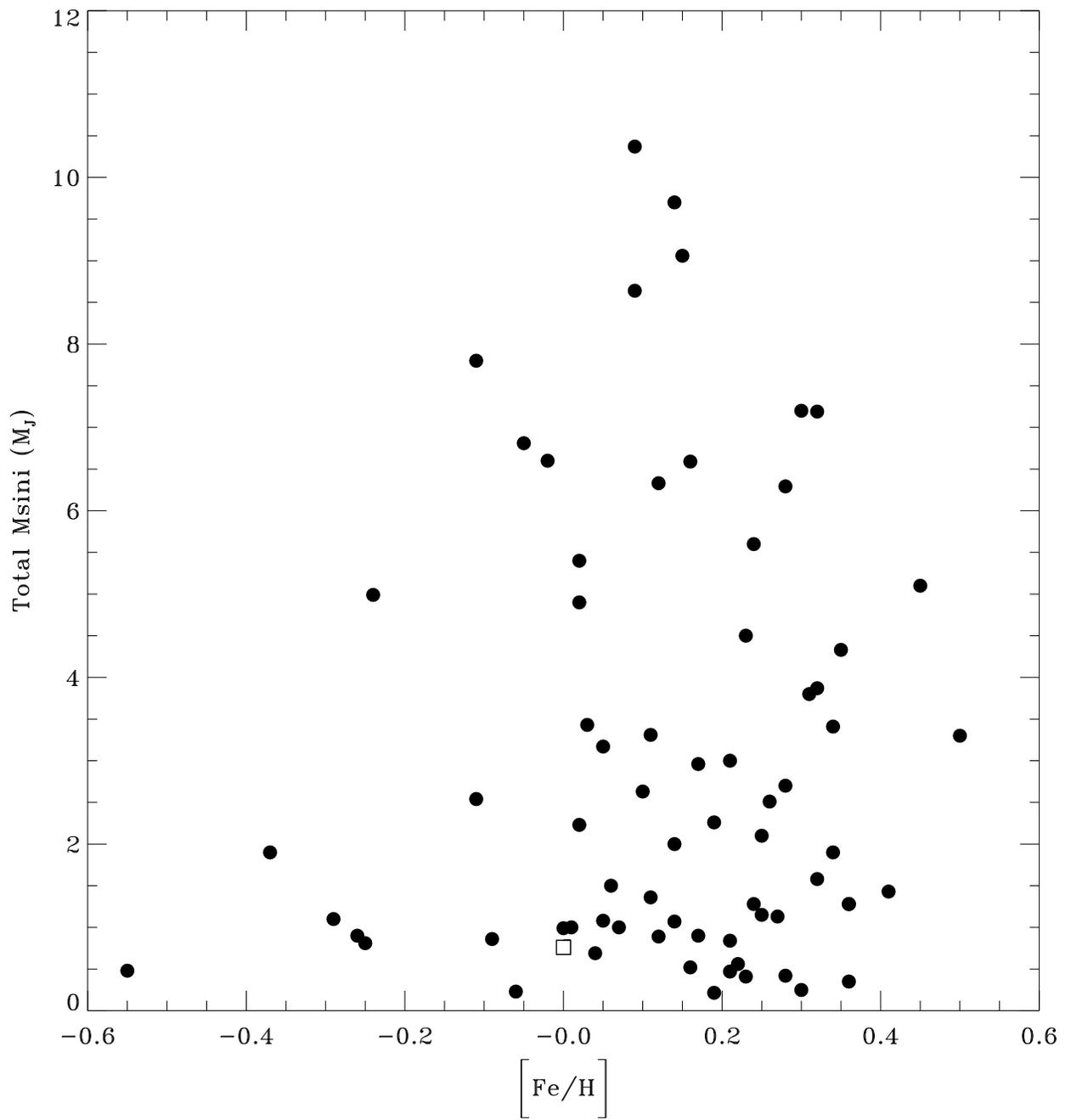
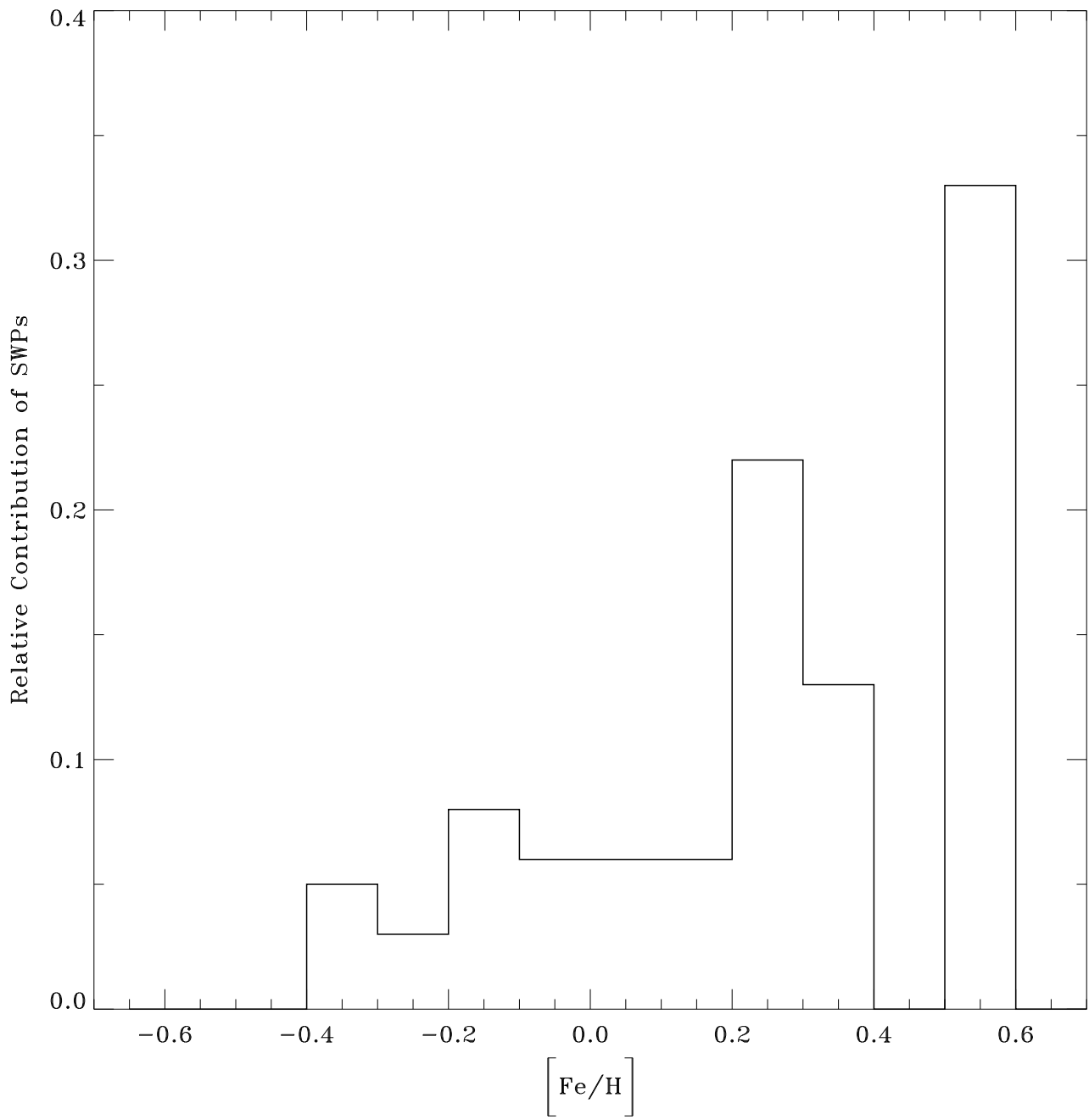
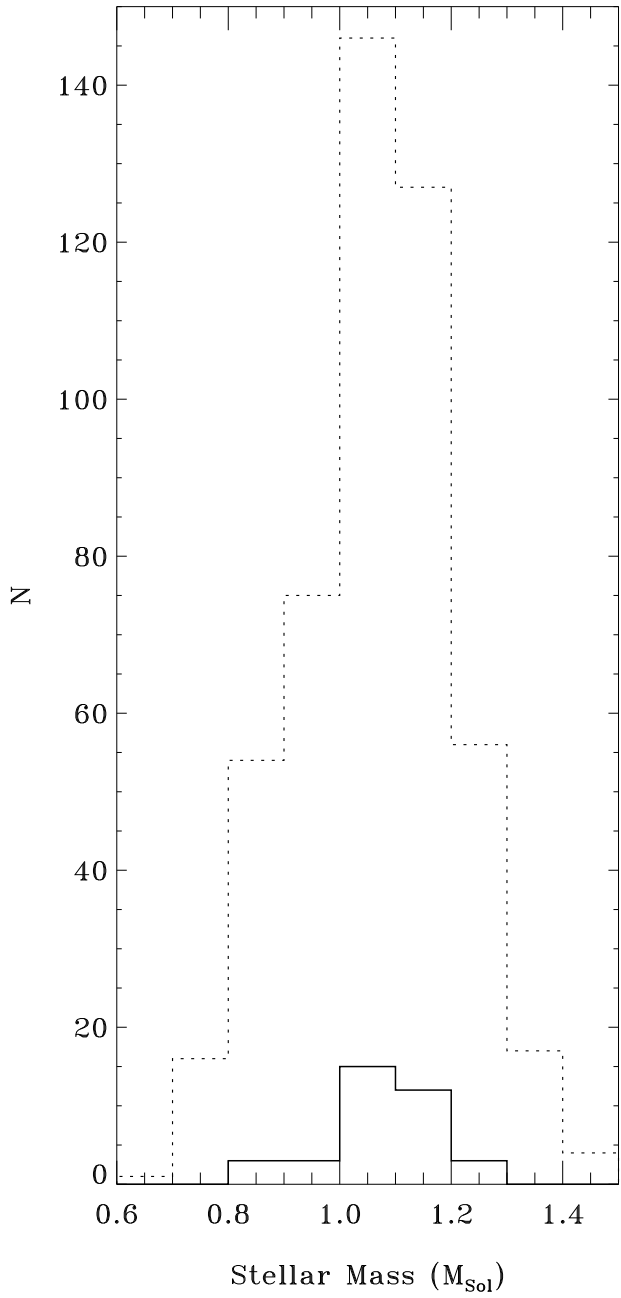
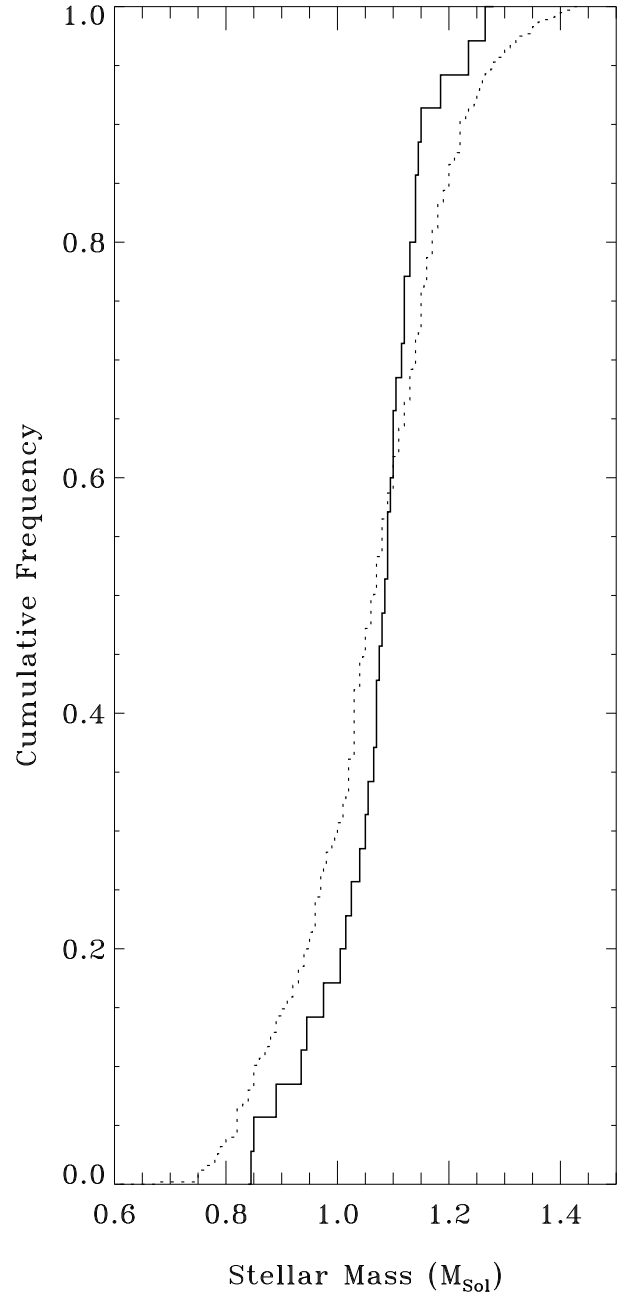
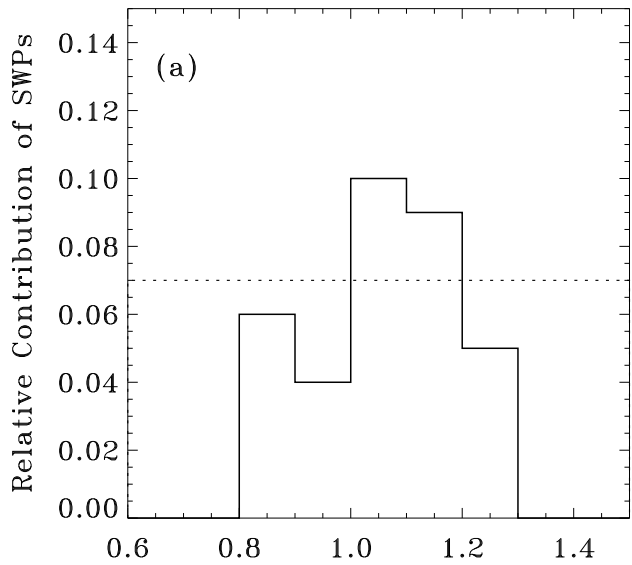
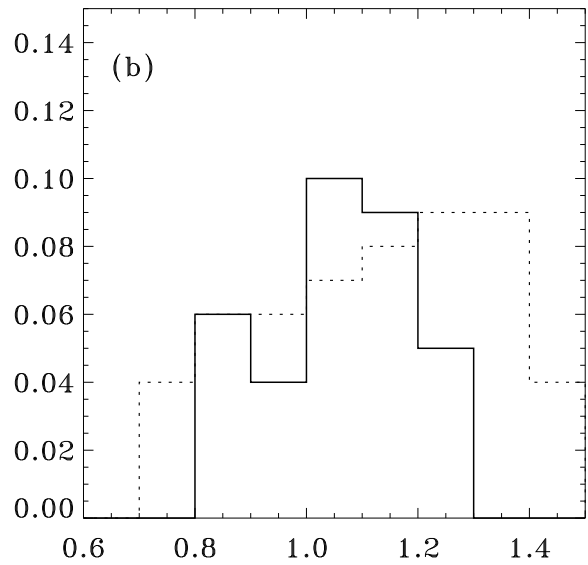
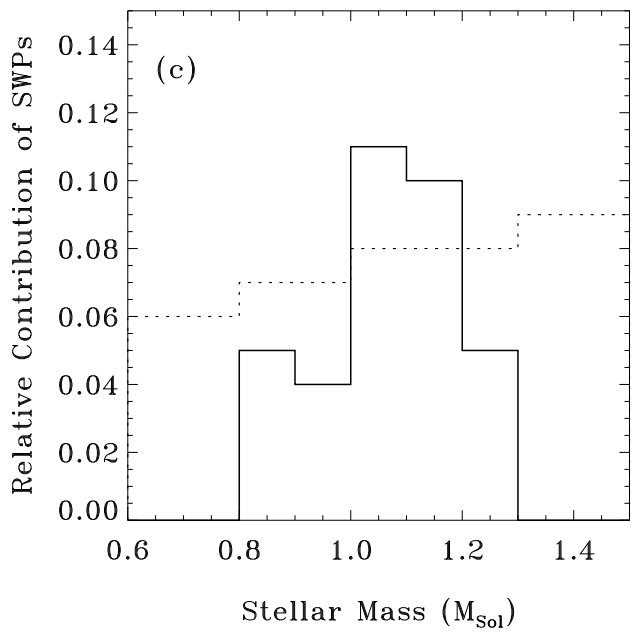
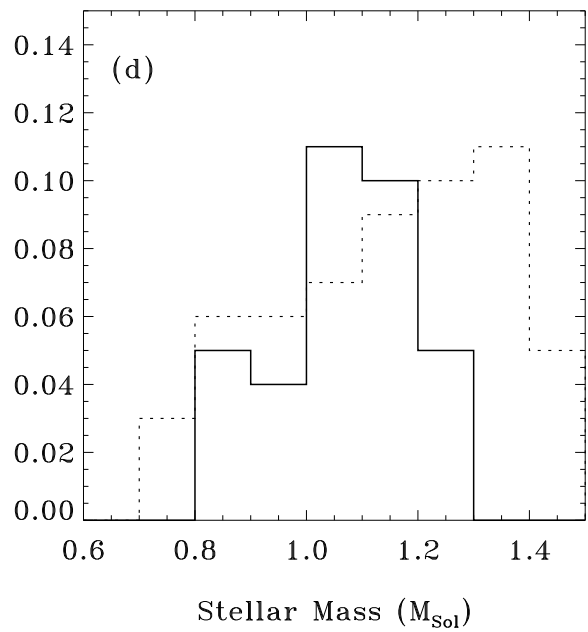
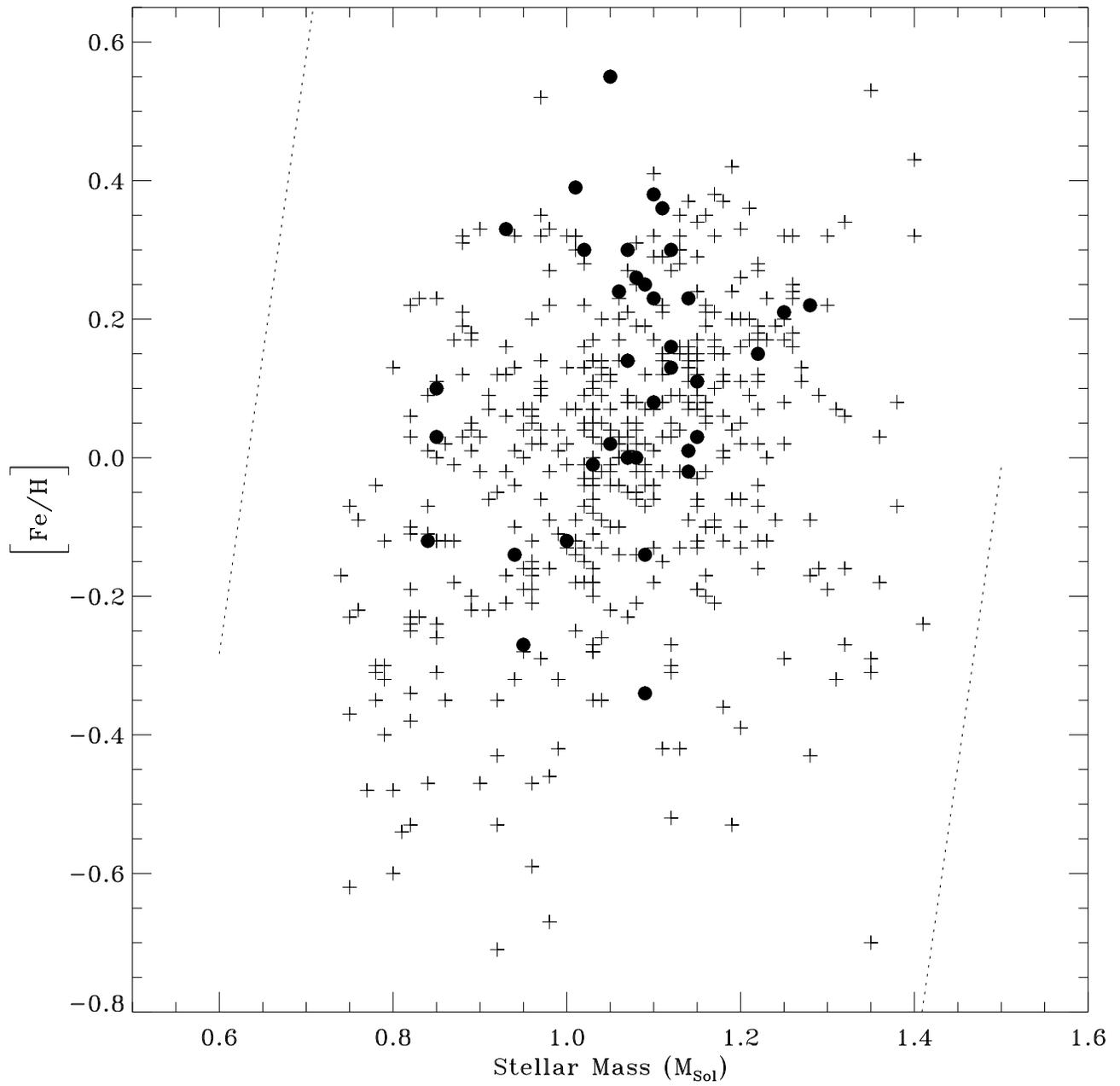
\end{document}